\documentclass[aps,12pt,amsmath,amssymb, tightenlines]{revtex4}
\usepackage{epsf,amsmath,graphicx, amssymb}
\usepackage{natbib}
\bibpunct{(}{)}{,}{a}{}{,}

\newcommand{\eq}[1]{\begin{equation}#1\end{equation}}

\renewcommand{\vec}[1]{\ensuremath{\mathbf{#1}}}
\newcommand{\D}{\ensuremath{\mathrm{d}}}
\newcommand{\Di}[1]{\ensuremath{\,\mathrm{d}#1\,}}
\newcommand{\ep}[1]{\ensuremath{\mathrm{e}^{#1}}}
\begin{document}

\title{Glass Transitions and Shear Thickening Suspension Rheology}
\author{C. B. Holmes}
\affiliation{School of Physics, The University of
  Edinburgh, JCMB, The King's Buildings, Edinburgh, EH9 3JZ, UK.}
\author{M. Fuchs}
\affiliation{Fachbereich Physik, Universit\"{a}t Konstanz, D-78457
  Konstanz, Germany.}
\author{M. E. Cates}
\email[Corresponding author. Electronic mail: ]{mec@ph.ed.ac.uk}
\affiliation{School of Physics, The University of
  Edinburgh, JCMB, The King's Buildings, Edinburgh, EH9 3JZ, UK.}
\author{P. Sollich}
\affiliation{Department of Mathematics, King's College, University of
  London, Strand, London, WC2R 2LS, UK.}

\date{\today}

\begin{abstract}
We introduce a class of simple models for shear thickening
and/ or `jamming' in  colloidal suspensions. These
are based on schematic mode coupling theory (MCT) of the glass
transition, having a memory term that depends on a density variable,
and on both the shear
stress and the shear rate. (Tensorial aspects of the rheology, such as
normal stresses, are ignored for simplicity.)
We calculate steady-state flow curves and
correlation functions. Depending
on model parameters, we find a range of rheological behaviours,
including  `S-shaped' flow curves, indicating discontinuous shear
thickening, and stress-induced transitions from a fluid to a
nonergodic (jammed) state, showing zero flow rate in an interval of
applied stress. The shear thickening and jamming scenarios that we
explore appear broadly consistent with experiments on dense colloids
close to the glass transition, despite the
fact that we ignore
hydrodynamic interactions. In particular, the jamming
transition we propose is conceptually quite different from various
hydrodynamic mechanisms of shear thickening in the literature,
although the latter might
remain pertinent at lower colloid densities.  Our jammed
state is a stress-induced glass, but its nonergodicity
transitions have an analytical
structure distinct from that of the conventional MCT glass transition.
\end{abstract}
\maketitle

\section{Introduction}
\label{intro}
Under flow, concentrated hard sphere suspensions can shear thicken: the measured viscosity
increases with the applied stress or flow rate [\cite{Frith, d'Haene, Laun, Bender, O'Brien+Mackay}]. The effect
can be pronounced, in some cases giving an order-of-magnitude jump in
the shear stress on exceeding a `critical' shear rate. In such
`discontinuous' shear thickening the flow often becomes erratic, suggesting
the continual formation and breakup of stress-supporting structures
[\cite{Frith}].  This idea is supported by recent
pressure-driven flow experiments on model colloids, in which
transient solid regions are directly visualised [\cite{HawPRL04}].
Computer simulations give a similar picture, with
simulations at infinite Peclet number 
suggesting that, in the absence of Brownian motion, a
static, load bearing jam may form [\cite{Melrose+Ball95}].
This is reminiscient of
jamming in granular materials, in which the solidity of the material
may result from the applied stress itself [\cite{Forcechains}].

The notion that strongly shear thickened states have the character of
an amorphous solid, or glass, is supported by recent experiments
[\cite{Bertrand2002}],
in which a droplet of shear thickening
material undergoes a fluid-solid transition upon shearing. 
In these experiments, a suspension (mean particle radius $a \approx 3 \mu$m) was
transformed into a solid state, simply by
stirring with a spatula. The jammed solid persists after
cessation of stirring, but is
refluidised by vibration. Also, it is
reported that introducing a droplet of the quiescent liquid
state adjacent to a jammed droplet (at the same concentration) results in the
two coalescing to form a larger fluid droplet. The description of
these phenomena reported in \cite{Bertrand2002} is brief, but
similar observations have
been made in suspensions of
zeolite particles  
and
hard-sphere colloids 
. (Note that the system of \cite{Bertrand2002}
is not hard-sphere like, and the reported phenomena are at relatively
modest concentrations. For hard spheres, the relevant concentration range lies
above 50\% by volume.) Thus there is some evidence of
two distinct states of a dense suspension, one fluid and one jammed.
These must presumably have different internal organization. Indeed,
the surface of jammed droplets appears dry or dull (becoming shiny
again only when refluidised by vibration or coalescence) suggesting
that partial emergence of particles is needed to sustain the jammed
state. This is consistent with the presence of large, static
deviatoric stresses within the bulk of a jammed droplet, balanced at
the surface by capillary forces.

Such experiments pose important challenges to the rheological
modelling of colloidal suspensions. In particular they suggest, though
they do not prove, that static shear stresses might be sustained
indefinitely in a bulk sample at rest in a rheometer -- even if the
same sample is fluid at lower stresses. (In this paper we address only
this bulk geometry; we hope to return to freestanding droplets in
future work.) On the other hand, shear thickening is commonly
attributed to `hydrodynamic clustering' [\cite{Melrose+Ball95, Brady,
    Ball+Melrose95, Bergenholtz}] 
-- that is, the formation of transient clusters of particles
in lubrication contact. This results in 
large hydrodynamic stresses in fast
flowing suspensions, and there is simulation and other evidence that these are the 
dominant stresses in the shear-thickened regime [\cite{Bender, Brady,
  PhungBradyBossis}]. However, some of these simulations are limited
to relatively modest volume fractions of colloid ($\phi \le 0.49$);
experiments at much higher concentrations
suggest instead a thermodynamic mechanism [\cite{O'Brien+Mackay}]. In
any case it seems highly unlikely that the `static jamming' observed
by \cite{Bertrand2002} is due to hydrodynamic interactions alone, since in the
persisting solid phase there is no macroscopic flow to provide those
interactions. Also, as far as we know, hydrodynamic models for shear
thickening offer no immediate explanation of nonmonotonic regions of
the flow curve. Such regions may be observed directly
[\cite{Frith,Bertrand2002}], and can lead (via well understood
shear-banding mechanisms [\cite{OlmstedReview}]) to
discontinuous shear thickening and accompanying hysteresis [\cite{Laun}]. 

Thus it seems clear that mechanisms other than pure
hydrodynamics are at work in sufficiently dense shear-thickening suspensions
[\cite{O'Brien+Mackay}]. Indeed, it has already been emphasised
[\cite{Melrose+Ball95,Ball+Melrose95}] 
that  deviations from pure lubrication forces can dominate
the physics of any hydrodynamically clustered state. (These deviations
could arise, for example, from deformation of stabilizing polymer
layers around the particles [\cite{Melrose+Ball95,Ball+Melrose95,
  MelroseVanVlietBall96, MelroseFaraday}].) 
Crucially, hydrodynamic theories neglect something else too: the colloidal glass transition [\cite{PuseyLesHouches}].

In quiescent
colloidal fluids, Brownian relaxation processes are often interpreted
in terms of particle `caging'. There is a
short-time relaxation, called $\beta$--relaxation, which is a many-body
process during which particles remain within
a cage of nearest neighbours. This is followed by a late-time
$\alpha$--relaxation, which corresponds to particles escaping their
cages. As the glass transition is approached, the cages become tighter, and both of these
relaxation times become longer. In
quiescent dense suspensions, it is this mechanism (rather than
simply the vanishing of mobility as interparticle gaps close to zero separation at random
close packing) that leads to the observed slow relaxations and large 
viscosities. Within the so-called mode coupling theory (MCT) of the
glass transition, the divergence of the $\alpha$--relaxation time (and
hence the viscosity) occurs at a colloid volume fraction $\phi_g$ around
52\% (although this value is very sensitive to the input static
structure factor). This is far below the random close packing density (64\%) which is
the natural location of any divergence in purely hydrodynamic
theories. In practice, this MCT value is too low, but not by much --
light scattering data on $\alpha$--relaxation locate the glass
transition at 58\% [\cite{vanMegenUnderwood}].

We contend that the slow relaxation of thermodynamic forces caused by particle 
caging remains important in sheared dense
suspensions, and is key to understanding their non-Newtonian
rheology. If this is true, the interesting physics should arise at
shear rates $\dot\gamma$ set by a renormalized Peclet number
$\dot\gamma\tau_\alpha$ involving the $\alpha$--relaxation time. In
contrast, hydrodynamic mechanisms are controlled, in the simplest
approximation,
by the bare Peclet number $\dot\gamma \tau_0$, with $\tau_0$ the
diffusive relaxation time of an isolated particle [\cite{Bender,
  PhungBradyBossis}]. 
Less naively, one could expect this to be renormalized
to $\dot\gamma\tau_1$ where $\tau_1$ is a lubrication-diffusion time
that diverges at random close packing [\cite{O'Brien+Mackay,
  FossBrady00}]. But as emphasised above, even $\tau_1$ is vastly shorter
than $\tau_\alpha$ in fluids close to the glass transition, and
infinitely shorter throughout the glass phase itself, where
$\tau_\alpha$ has diverged.

While in practice the hydrodynamic and thermodynamic forces in very
dense colloids may be interdependent, this separation of time scales
makes it legitimate to explore the possible connection between the
glass transition and shear thickening in its own right, ignoring
hydrodynamic interactions in the first instance. 
From this viewpoint, the jammed state found experimentally
[\cite{Bertrand2002}] is a candidate for a stress-induced, anisotropic 
colloidal glass. Such a state could arise if 
the applied stress alters the material's  structure,
leading to an increased number of close contacts,
hindering diffusion and promoting dynamical
arrest through the tightening of the cage around each particle.

In this picture, as in a conventional glass, solidity is due to
thermodynamic forces (purely entropic for hard spheres): deformation of the material results in
a free energy penalty which
appears macroscopically as elasticity. An alternative is that Brownian motion
is sufficiently weak that these entropic forces are irrelevant,
leaving interparticle forces alone to resist deformation: 
in hard spheres, stress applied to the jammed solid is then transmitted by
direct contacts. In this case, the jamming more closely resembles that of a
granular medium [\cite{Forcechains}]. (In fact it has been suggested
by \cite{LiuNagel} that, even in this non-Brownian case, arrest is
related to a glass transition; we do not pursue this here.) 
In a jammed suspension, it is not clear which mechanism dominates; the answer might depend strongly on particle
size. Nor is it clear under what conditions static jamming may be achieved:
sufficient Brownian motion might give all or most jammed states a
finite lifetime. Note that, in contrast to some authors (for example
\cite{LiuNagel}), in this work we use the word `jamming'
to mean an arrest transition caused specifically by imposed
stress rather than other causes, such as attractive interactions.

Our modelling strategy, which was outlined with selected results in
\cite{HolmesEPL}, 
neglects hydrodynamics, and assumes a 
 dynamical glass transition picture rather than one based on
direct mechanical contact
between particles [\cite{Forcechains}]. It builds
upon a schematic version of MCT, and is guided by more formal (and
less schematic) MCT work on the role of shear in the glass transition
[\cite{Fuchs+CatesPRL,Fuchs+CatesFaraday}]. The latter formal approach
currently gives only shear-thinning behaviour, or strain-induced
fluidisation. To this we add a new feature in our schematic models:
stress-induced arrest. (Capturing the same physics, beyond the
schematic level, remains an open issue for future development of the
more formal theory [\cite{CatesPoincare}].)
In combination, these two features suggest
various shear thickening and jamming scenarios, some of which appear
broadly consistent with the experimental features noted above. 

In some but not all cases, these scenarios are similar to
ones found before from a quite different approach to glassy rheology,
originally intended for `soft glasses' (foams, dense emulsions) rather than
colloidal suspensions. In particular, \cite{Head01,
  Head02EPL} incorporated
an ad-hoc local jamming effect into the `soft glassy rheology model'
(SGR) [\cite{SollichEtAl97, Fielding00, CatesSollich04}] thereby converting its shear-thinning behaviour 
into a shear thickening one. Our work on schematic MCT, though equally ad-hoc at 
this stage, has the advantage that one day it might be underpinned by full MCT-type
calculations, whose form might well be guided by comparing our different schematic model variants 
with experimental data. (A drawback is that, unlike SGR-based work, we can
only address steady state properties and not dynamical ones.) Some of
the flow curves we obtain are also related to those found by
\cite{Hess} although their work did not attempt to relate shear 
thickening to glass transitions.

The rest of this paper is organized as follows. Section \ref{MCT} provides a
brief introduction to MCT; in Sec. \ref{model}, we
formulate our new class of schematic models. Analytical and numerical methods
are described in Sec. \ref{solution} and the results of these presented in
Secs. \ref{Results} and \ref{variations}. Sec. \ref{Discuss} contains
a discussion of our results in the context of experiment, and we
conclude in Sec. \ref{conclusions}.

\section{Mode Coupling Theory}\label{MCT}
For a comprehensive description of MCT, see \cite{Goetze}. Here, we
outline only those aspects pertinent to the current work. 
The central quantities of the theory are the Fourier-space density fluctuations at
wavevector ${\mathbf q}$,
   $\,\delta\rho({\mathbf q},t)$. The correlators of these quantities,
$\phi_{\mathbf q}(t)\equiv\langle\delta\rho({\mathbf
  q},t)\delta\rho(-{\mathbf q},0)\rangle/\langle|\delta\rho({\mathbf
  q})|^2\rangle$ may be measured in scattering experiments [\cite{PuseyLesHouches}], and describe the system's dynamics. In a liquid, the
system is ergodic and  $\phi_{\mathbf q}(t)$ decays to zero with time
for all ${\bf q}$. In a glass it does not: $\lim_{t\to\infty}
\phi_{\bf q}(t) =f_{\bf q}>0$, where the nonergodicity
parameters $f_{\bf q}$ characterise the arrest.  A finite $f_{\bf
  q}$ implies inability to relax on
a lengthscale $\sim 2\pi/q$. 
(In general, all ${f_{\bf q}}$ become nonzero at the same transition point; once one Fourier component of density is frozen, the others see this as a random potential and acquire non-decaying mean values.)

Within MCT, equations of motion are found for the correlators $\phi_{\mathbf
  q}(t)$. Assuming overdamped local motion appropriate to colloids,
and dropping $\mathbf{q}$-subscripts, the result is
[\cite{GoetzeEssentials}]:
\begin{equation}
\phi(t)+\tau_o\dot{\phi}(t)+\int_0^tm(t-t^\prime)\dot{\phi}(t^\prime)\,\mathrm{d}t^\prime=0,\label{EOM}
\end{equation}
where $\tau_o$ is a timescale set by the microscopic dynamics. Here
$m(t-t^\prime)$ is the {\it memory function}, and describes a retarded
friction which, in the colloidal glass transition, arises by
caging of a particle by its neighbours. In MCT, the memory function is
found approximately by integrating (over wavevectors) a quadratic
product of correlators, with coupling constants that depend solely on the
static structure factor $S(q) = (1/N)\langle|\delta\rho({\mathbf
  q})|^2\rangle$ of the system. The $\tau_o$ term represents instantaneous solvent friction (with many body hydrodynamics omitted)
and gives exponential relaxation of correlations in
dilute colloids, in which memory effects are negligible.
Note that the nature of the interparticle interactions enters through
$S(q)$; MCT is not limited to hard-sphere colloids.

In MCT, the zero shear viscosity diverges with the $\alpha$--relaxation
time of the density correlations. An infinitesimal stress
applied to a colloidal glass state gives zero steady-state shear rate $\dot\gamma$: glasses are solid, although creep (sublinear in time) is not excluded.
The extension of MCT to finite $\dot\gamma$ requires care, as Eq. \ref{EOM}
relies upon the fluctuation dissipation theorem
(FDT) [\cite{BouchaudEtAl}] which need not hold
under shear [\cite{BarratBerthierFDT}]. However, extended forms
of MCT describing shear have been formulated, either with the
assumption of an FDT--like condition
[\cite{miyazaki}], or without it
[\cite{Fuchs+CatesPRL,Fuchs+CatesFaraday}]. These approaches lead
 to equations of motion for
correlators like $\phi_{\bf q}$ (redefined to account for simple
advection) which formally resemble those of the
unsheared theory.  Solution of these equations,
performed via MCT--like approximations, leads to the
conclusion  that structural relaxation always speeds up in a flowing
system: $\tau_\alpha(\dot\gamma) \le \tau_\alpha(0)$. (Analogous behaviour is seen in mean field spin models of
driven glasses [\cite{BBK}].) Physically, this is because small scale 
diffusive motions are
magnified by advection, facilitating break up of the cage on a
timescale set by $1/\dot\gamma$. In
$\mathbf{q}$--space, the important fluctuations, which have
wavevectors near the peak of $S(q)$, are advected on this timescale to
higher $q$ where they decay rapidly. Even those fluctuations which are
not directly advected become ergodic [\cite{Fuchs+CatesPRL,Fuchs+CatesFaraday}].

So far then, these microscopic extensions of MCT to sheared systems do not
predict jamming under shear stress. This may be because shear thickening
requires hard-core particle interactions which are not fully captured 
by the harmonic approximation to the thermodynamic forces conventionally 
made in MCT [\cite{CatesPoincare}].  
The latter is tantamount to an expansion of the free energy to second order in density fluctuations; in order to see the jamming effect, one might have to go beyond this order. 
We do not attempt this formidable task
here: rather, we adopt a simpler approach, beginning from `schematic
models' of the MCT glass transition. These stripped-down models
represent a gross simplification of the full theory and yet manage to
capture many of the key features of the quiescent glass transition [\cite{Goetze}]. In them, one considers a single
correlator $\phi(t)$ (which may be considered to represent fluctuations at some
typical wavevector), rather than the infinite
set $\phi_\mathbf{q}(t)$. The memory function is then written as a
polynomial of this correlator, with coefficients (coupling constants)
that encode the interactions between particles (just as $S(q)$ encodes these in full MCT). Larger coupling
constants correspond to higher colloid densities and/or lower temperatures. 

Of interest here are two schematic models, known as the
F2-- and F12--models, defined by a shared equation (Eq. \ref{EOM})
for the correlation function $\phi(t)$,
and by the memory functions
\begin{eqnarray}
m(t)&=&v_2 \phi^2(t)\;\;\;\;\;\;\;\;\;\;\;\;\;\;\;\;\;\;\;\,\mbox{(F2 model)}\label{F2}\\
m(t)&=&v_1\phi(t)+v_2\phi^2(t) \;\;\;\;\;\;\mbox{(F12 model)}\label{F12}
\end{eqnarray} 
Each model exhibits a glass transition with increasing $v_n$; for the F2 model, this
occurs at $v_2^c=4$ whilst in the F12 model (which has $v_2 >1$ [\cite{Goetze}]), there is a locus of
transition points given by $v_1^c=v_2\left(\sqrt{4/v_2}-1\right)$. 
As these transitions are crossed, the single nonergodicity parameter
$f\equiv\phi(\infty)$
jumps discontinuously, from zero on the liquid side of the transition,
to a finite value $f_c$ on the glass side. 
On further increasing the coupling, there is a
nonanalytic (square root) increase in $f$ beyond $f_c$. 
This behaviour of $f$ is faithful to the full MCT in which a similar
discontinuity and square root law are found for the $f_{\bf q}$.

\section{Rheological Models}\label{model}
To incorporate shearing into the schematic MCT models we need
to address two distinct effects. Firstly flow
erodes memory: particles with nonzero separation in the flow
gradient direction become separated in the flow direction on
times $t\sim1/\dot\gamma$. Thus, diffusive motion
is accentuated by the flow, with the result that cages are broken and
correlations decay to zero. 
The second effect is that stress (as distinct from flow) can alter
the local caging of particles such as to promote arrest. 
Within a schematic theory, the latter requires us to postulate a stress dependence of the coupling constants.
Perhaps the simplest idea is to 
assume a Taylor expansion of the coupling
constants $v_n$ ($n=1,2$) in Eqs. \ref{F2},\ref{F12}, that is,
$v_n(\sigma) = v_n+\sum_k\alpha_k\sigma^k$.
One could then neglect odd powers on the grounds of symmetry. 
However, this approach is limited to the region around $\sigma=0$, whereas
we are interested in nonzero stresses (in particular,
values close to the yield stress of an incipient glass). Also, it
turns out that the topology of flow curves depends strongly on the
form of $v_n(\sigma)$ at large $\sigma$, rather than that near the
origin. Hence, although we choose mainly simple power laws for
$v_n(\sigma)$, these are not to be viewed as a Taylor expansion around
$\sigma=0$. In particular, a linear dependence $v_n(\sigma)= v_n+\alpha\sigma$, is
not excluded by symmetry. Here, and from now on, $\sigma$ denotes the
{\em absolute value} of the shear stress. The various model forms
chosen for $v_n(\sigma)$ are specified below in
Eqs. \ref{MainMemoryFn}--\ref{ModelV}.

To allow for the flow-induced memory loss, the memory function
should also be attenuated by a strain dependent factor
[\cite{Fuchs+CatesPRL,Fuchs+CatesFaraday}. This suggests that we write
\begin{equation}\label{memory}
m(t)=\sum_n v_n(\sigma) \,\phi^n(t)f(\dot\gamma t),
\end{equation}
where the function $f(\dot\gamma t)$ is monotonically
decreasing. Microscopic theory [\cite{Fuchs+CatesPRL,
Fuchs+CatesFaraday} suggests a specific form: $f(\dot\gamma
t)=(1+\dot\gamma^2t^2)^{-1}$. But for algebraic convenience, we shall mainly consider an exponential
form $f(\dot\gamma t)=\exp{(-\dot\gamma t)}$.

In Eq. \ref{memory} for the memory function, the effect of shearing appears in two
quite separate ways: via the applied stress $\sigma$, and with the resulting
(steady state) shear rate $\dot\gamma$. These two quantities 
are related through the viscosity \eq{\eta\equiv
\sigma/\dot\gamma.} In order to close the
model, we now need a prescription for the viscosity. In linear response
theory, the shear viscosity is expressed as the time
integral of a stress correlator [\cite{HansenMcDonald}]. 
In our simplified model, we have but one correlator, and so we choose
schematically to identify the viscosity as 
\begin{equation}\label{ViscosityPrescription}
\eta=\int_0^\infty \phi(t)\,\mathrm{d}t=\tau,
\end{equation}
where the appropriate elastic modulus has been set equal to unity,
rendering the stress dimensionless. Eq. \ref{ViscosityPrescription} is
somewhat {\it ad hoc}, since we are not studying the 
linear response regime, but experiment suggests
a close relationship between viscosity and relaxation times
[\cite{MeekerEtAl97}]. Eq. \ref{ViscosityPrescription} means that
dynamical arrest (nonzero $f$) implies a divergent viscosity, and hence
zero shear rate, in steady state. 
 This is fully consistent with arguments made above
that finite $\dot\gamma$ precludes arrest. Note that
 equating $\eta$ to an integral of
$\phi^2(t)$ (say) would give very similar results.

Eqs. \ref{EOM}, \ref{memory} and \ref{ViscosityPrescription}
are shared by all the models in the class we study here. Different
members of the class have different choices of $v_n(\sigma)$ and
$f(\dot\gamma t)$ in Eq. \ref{memory}. Among the variants we shall address are
\begin{equation}\label{MainMemoryFn}
m(t)=(v_2+\alpha\sigma)\phi^2(t)\exp{(-\dot\gamma t)}\;\;\;\;\;\;
\mbox{\rm Model I}
\end{equation} 
\begin{equation}\label{ModelII}
m(t)=(v_2+\alpha\sigma^2)\phi^2(t)\exp{(-\dot\gamma t)}\;\;\;\;\;\;
\mbox{\rm Model II}
\end{equation} These variants are
amenable to an analytic approach, which we present in the next Section.
We also consider a Model III which interpolates between
Model II at small stresses and Model I at large ones; this has 
\begin{equation}\label{ModelIII}
m(t)=(v_2+\alpha\inf[\sigma,\sigma^2])\phi^2(t)\exp{(-\dot\gamma t)}\;\;\;\;\;\;
\mbox{\rm Model III}
\end{equation} 
The other variations which we shall study are:
\begin{equation}\label{ModelIV}
m(t)=(v_2+\alpha\sigma)\phi^2(t)/(1+\dot\gamma^2t^2)\;\;\;\;\;\;
\mbox{\rm Model IV}
\end{equation} 
\begin{equation}\label{ModelV}
m(t)=\left[(v_1+\alpha_1\sigma)\phi(t)+(v_2+\alpha_2\sigma)\phi^2(t)\right] \exp{(-\dot\gamma t)}\;\;\;\;\;\;
\mbox{\rm Model V}
\end{equation} 
Model IV differs from model I only in that the choice of $f(\dot\gamma t)$ is closer to the form expected from the full MCT [\cite{Fuchs+CatesPRL,Fuchs+CatesFaraday}].
Unfortunately this limits analytic progress, but we present below numerical
results for this case. Model V departs from variants I-IV by being based on the
F12 model rather than F2; similar remarks apply. 

In each of these model variants, there are control parameters such as
$v_n$ and $\alpha$: these are expected
to have nontrivial dependence on both the concentration of colloids and their interactions.
Since in practice a range of different shear-thickening scenarios 
are observed for different materials, an attempt at a parameter-free theory would, at this
stage, be misguided. We choose $\alpha \ge 0$ in this paper.

\section{Solving the Models}\label{solution}
\subsection{Analytical Developments}\label{analytic}    
Our analytical work is based on the F2 model and hence restricted to Models I-IV, for which $v_1 = 0$. The methods are adapted from standard techniques used in
solution of the MCT equations of motion (see, {\it eg}
[\cite{Goetze}]). We begin by defining a Laplace transform
\begin{equation}
f(z)\equiv i\int_0^\infty f(t) \exp(i z
t)\,\mathrm{d}t;\, \,\, \Im(z)>0.
\end{equation}
Then, Laplace transforming Eq. \ref{EOM}, we obtain
\begin{equation}
\phi(z)=\frac{-1}{z-\frac{1}{i\tau_o+m(z)}}.\label{Laplace1}
\end{equation}
The dynamical slowing down associated with the glass transition is
contained in $m(z)$. In contrast, the instantaneous friction term
remains constant, and so, close to the transition, 
$|\tau_o|\ll|m(z)|$ for low frequencies. For clarity in
the following, we label the position in parameter space by the
vector $\mathbf{v}\equiv (v_2, \alpha, \sigma)$ so that, {\it eg}, $\phi_\mathbf{v}(z)$ is the
frequency-space correlator at frequency $z$ for parameters given by
$\mathbf{v}$. (Likewise, $\dot\gamma_\mathbf{v}$ is the shear rate
for the given parameter values.)
At low frequencies then, we neglect the instantaneous
friction term to give 
\begin{equation}
\frac{z\phi_\mathbf{v}(z)}{1+z\phi_\mathbf{v}(z)}=zm_\mathbf{v}(z) \label{LaplaceEOM}.
\end{equation}

We now split the correlator into the sum of its late time limit and a
(time-dependent) remainder, writing  $\phi_\mathbf{v}(t)= f_\mathbf{v}+g_\mathbf{v}(t)$. 
 Now we assume that, for suitable values of the relevant
parameters, our model admits nonergodic solutions. 
In a nonergodic state the relaxation time $\tau$ is
divergent and so $\dot\gamma=0$ for any finite
stress. Therefore the zero-frequency limit of
Eq. \ref{LaplaceEOM} gives
\begin{equation}
\frac{f_{\mathbf v}}{1-f_{\mathbf v}}=Vf_{\mathbf v}^2\label{bifurcation},
\end{equation}
where $V\equiv v_2(\sigma)$. Here, we have used the fact that, since  $g_{\mathbf{v}}(t)$ tends to zero at 
late times, terms of order $g_\vec{v}(t)$ or $g^2_\vec{v}(t)$ give rise to
singularities weaker than $1/z$ as $z\rightarrow 0$. It
has been shown [\cite{GoetzeSjogrenJMathAnal}] that the nonergodicity parameter ($\lim_{t\to\infty}
\phi_\vec{v}(t)$) is the largest of the
solutions to Eq. \ref{bifurcation}. 

This result fixes the
nonergodicity parameter of an arrested state with coupling $V$: there are
nonzero solutions for $f_\mathbf{v}$ provided $V\ge 4$, so that
a glass becomes possible beyond this value of the coupling. 
However, this does not imply that we necessarily have a glass
transition at $V=4$, since we have calculated this by assuming
arrest, which requires $\dot\gamma = 0$. (Without shear, our 
model is simply an F2 model 
for which the result $V_c = 4$ is well known.) 
In Sec. \ref{MCT}, we stated that passing through a static MCT glass transition entails
a discontinuity in the nonergodicity parameter $f$ followed by a
nonanalytic variation as the coupling constants increase. 
Solving Eq. \ref{bifurcation} shows that such behaviour will arise in Models I-IV only for an arrest transition occurring {\it at} the F2
value of $V=4$. We shall return to this point later.

\subsubsection{Asymptotics for correlator and memory function}
We now consider the behaviour of  
correlation functions close to a 
transition at (say) $\vec{v}_c$, corresponding
to some $V>4$.
Near such a transition,  
$|\mathbf{v}-\mathbf{v}_c|$ is a small quantity. To identify other quantities which may be treated perturbatively, 
we recall static MCT behaviour: nonergodic solutions to
Eq. \ref{EOM} show a decay, at the $\beta$--timescale
$\tau_\beta$, onto a plateau, with
$\phi_\vec{v}(t>\tau_\beta)\approx f_\vec{v}$. 
For $V>4$, on timescales for which $\dot\gamma_\vec{v} t \ll 1$, the
difference $\phi_\mathbf{v}(t) - \phi_{\mathbf{v}_c}(t)$ is also a small
quantity.
Provided $\dot\gamma_\vec{v}$ vanishes smoothly at the jamming transition,
we can ensure that this condition holds at and beyond the timescale $\tau_\beta$ 
on which $\phi_{\mathbf{v}_c}(t)$ reaches its long time limit.
That is, we choose $|\mathbf{v}-\mathbf{v}_c|$ such that
$\dot\gamma\tau_\beta\ll 1$. (Our ability to do this depends on the fact that $\tau_\beta$ is
finite, unlike $\tau_\alpha$, whenever $V>4$.) Thus we have identified a time regime where $g_\mathbf{v}(t)\equiv \phi_\mathbf{{v}}(t)-f_{\mathbf{v}_c}$ is a small quantity. 
In the frequency domain, this gives $z\phi_\mathbf{v}(z)=-f_{\mathbf{v}_c} + zg_\mathbf{v}(z)$.
In line with standard MCT techniques [\cite{Goetze}], we now treat
$g_\mathbf{v}(z)$ perturbatively in the Laplace transformed equation of
motion, Eq. \ref{LaplaceEOM}. Expanding the LHS in powers of $zg_\mathbf{v}(z)$, we find
\begin{equation}
\frac{z\phi_\mathbf{v}(z)}{1+z\phi_\mathbf{v}(z)}
=\frac{-f_{\mathbf{v}_c}}{1-f_{\mathbf{v}_c}}+\frac{zg_\mathbf{v}(z)}{(1-f_{\mathbf{v}_c})^2}-\frac{z^2g_\mathbf{v}^2(z)}{(1-f_{\mathbf{v}_c})^3}+\mathcal{O}(z^3g_\mathbf{v}^3(z)).
\end{equation}
We now wish to approximate the RHS of Eq. \ref{LaplaceEOM} in the same
regime of frequency. By considering the memory function in the time domain,
$\dot\gamma_\vec{v} t$ and $g_{\mathbf{v}}(t)$  can be treated
perturbatively. So can the distance  from the
transition in parameter space: we write 
$V=V_c+\epsilon$ where $V_c$ is the value of $V\equiv v_2(\sigma)$ at
$\mathbf{v}_c$.

To make further analytic progress we now restrict the form of the function $f(\dot\gamma t)$ in Eq.\ref{memory} to be exponential. 
(This excludes Model IV; we have already excluded Model V.) We find 
the memory function then to obey
\begin{equation}m_\vec{v}(t)=(V_c+\epsilon)
\left(f_{\mathbf{v}_c}^2+2f_{\mathbf{v}_c}
g_\mathbf{v}(t)+g_\mathbf{v}^2(t)\right)\left(1-\dot\gamma_\vec{v}
t+\mathcal{O}(\dot\gamma_\vec{v}^2 t^2)\right).
\end{equation}
In the time regime of interest, terms $\mathcal{O}(g_\vec{v}(t)\dot\gamma_\vec{v} t)$ and $\mathcal{O}\left(g^2_\vec{v}(t)\right)$ are negligible with
respect to terms $\mathcal{O}\left(g_\vec{v}(t)\right)$. Additionally, terms
$\mathcal{O}\left(\epsilon g_\vec{v}(t)\right)$ and $\mathcal{O}\left(\epsilon
\dot\gamma_\vec{v}t\right)$ may be neglected with respect to terms
$\mathcal{O}(\epsilon)$. 
The RHS of Eq. \ref{LaplaceEOM} then becomes 
\eq{zm_\vec{v}(z)\approx -V f_{\mathbf{v}_c}^2 + 2f_{\mathbf{v}_c}V_czg_\mathbf{v}(z) +V_cf_{\mathbf{v}_c}^2\dot\gamma_\vec{v}i/z. 
}
Using Eq. \ref{bifurcation}, we find
that close to the transition Eq. \ref{LaplaceEOM} becomes 
\begin{equation}
zg_{\mathbf{v}_c}(z)\left(\frac{1}{(1-f_{\mathbf{v}_c})^2}-2f_{\mathbf{v}_c} V_{c}\right) -V_{c}f_{\mathbf{v}_c}^2\dot\gamma_\vec{v}i/z
+f_{\mathbf{v}_c}^2\epsilon=0.
\end{equation}
Inverting the Laplace transform, we find  the correlator in
the time domain to be
\begin{equation}\label{alpha-decay}
\phi_\vec{v}(t)\approx f_{\mathbf{v}_c}\left(1+\frac{\epsilon
f_{\mathbf{v}_c}}{\frac{1}{(1-f_{\mathbf{v}_c})^2}-2f_{\mathbf{v}_c}V_c}\right)\left(1-\frac{V_cf_{\mathbf{v}_c}\dot\gamma_\vec{v}
t}{\frac{1}{(1-f_{\mathbf{v}_c})^2}-2f_{\mathbf{v}_c}V_c+\epsilon f_{\mathbf{v}_c}}\right),
\end{equation}
or, expressed more succinctly,
\eq{\phi_\vec{v}(t)=f\left(1-t/\tilde\tau\right)\label{dynamics}}
with nonergodicity parameter $f$ and a relaxation time $\tilde\tau$ defined appropriately. 

Eq. \ref{dynamics} is {\em consistent} with an exponential decay from the plateau at late times, $\phi\sim f\exp[-t/\tilde \tau]$, a result also found numerically in Sec. \ref{Results} below.
In what follows we further assume that this is the {\em correct}
late-time form for the correlator; this allows us to find explicitly the locus of arrest transitions arising in Models I-III.

We begin by evaluating Eq. \ref{Laplace1} at $z=0$ to deduce
the existence of a terminal relaxation time
\begin{equation}
\tau=\tau_o-im(0).\label{tau equation}
\end{equation}
As we approach a jamming transition at finite stress, the
relaxation time $\tilde \tau$ defined in Eq. \ref{dynamics} must diverge (since the
shear rate vanishes at the transition point). We continue to
assume that, at ${\bf v}_c$, $V>4$ in which case the 
$\tau_\beta$ remains finite. Therefore, approaching the
transition, the terminal time $\tau$ is primarily set by the
late-time exponential relaxation of $\phi_{\bf v}(t)$. Accordingly,
Eq. \ref{tau equation} becomes
\begin{eqnarray}
\tau&=&\tau_o+V\int_0^\infty\exp(-\dot\gamma
t)\phi^2(t)\,\mathrm{d}t\label{tau equation2}\\ 
&\approx&V\int_0^\infty\exp(-\dot\gamma
t)f^2\exp(-2t/\tilde\tau)\,\mathrm{d}t,
\label{tauagain}\end{eqnarray}
where we have approximated the correlator by $\phi(t)=f\exp(-t/\tilde\tau)$ 
over the whole temporal range (neglecting deviations at $t\lesssim \tau_\beta$), and dropped a (non-divergent)
contribution from the regular part of the memory function. By definition, $\tau=\int_0^\infty
\phi(t)\Di t$, and so  we
require $\int_0^\infty f\exp(-t/\tilde\tau)=\tau$, giving
$\tilde\tau=\tau/f$. 

We now write Eq. \ref{tauagain} as an iteration: the $nth$ approximation to the
relaxation time $\tau$ is calculated using the $(n-1)th$ value to
determine the shear rate. Thus 
\eq{\tau^{(1)}\approx \int_0^\infty Vf^2 \ep{-\sigma
t/\tau(0)}\ep{-2ft/\tau(1)}\D t,}
which we can solve to give the ratio of $\tau^{(1)}$ to $\tau^{(0)}$:

\eq{\frac{\tau^{(1)}}{\tau^{(0)}}=\frac{f}{\sigma}\left(Vf-2\right).\label{ratio}}
Crucially, this ratio is independent of
$\tau^{(0)}$: thus, no matter how large the initial guess, the
iteration leads to a larger value if the RHS is greater than
unity. Where this is the case, 
the relaxation time is divergent. Thus, setting the RHS in
Eq. \ref{ratio} to unity locates the arrest transition. Recalling the
definition $V\equiv v_2(\sigma)$, this lies
at a stress $\sigma_c$ which obeys
\eq{f_c\left[v_2(\sigma_c)f_c-2\right]-\sigma_c=0,\label{transitions}}
where $f_c$ is given by the largest solution of
$f_c/(1-f_c)=v_2(\sigma_c)f_{c}^2$ (from
Eq. \ref{bifurcation}).

Except for the special case $\sigma_c=0$, these transition points do not coincide
with the F2 transition at $V = 4$, and have a different character to MCT
transitions in general. From a rheological point of view, the transitions
give values of
the stress for which the shear rate first becomes zero: these are jamming
transitions. Their locations 
according to Eq. \ref{transitions} will later be checked against the
numerical solutions for the full flow curves.

\subsubsection{Large stress limit}\label{highstress}
We turn now to investigate the behaviour in the limit of large stress. We
begin by considering cases in which the relaxation time diverges in that limit: in this case, Eq. \ref{ratio} implies
\eq{\lim_{\sigma\to\infty} \alpha f^2 v_2(\sigma)/\sigma\ge 1. \label{largeStressTrans}}  The implications of this result depend upon
the model: if $v_2(\sigma)$ increases sublinearly at large $\sigma$,
this criterion will not be satisfied for any finite $\alpha$ and so
fluid-like behaviour is always recovered in the limit of large stress. 
For super-linear $v_2(\sigma)$, as arises in Model II, this condition places no
restriction upon $\alpha$ beyond $\alpha> 0$. Hence Model II 
always has zero shear rate at high stresses: the jammed state remains solid indefinitely. 

The most interesting case is when $v_2(\sigma)$ increases linearly at large $\sigma$, which applies in both Model I and Model III. Here Eq. \ref{largeStressTrans} can only be
satisfied if $\alpha\ge 1$, since in a jammed state the nonergodicity
parameter $f$ will approach unity in the limit of large stress (see
Eq. \ref{bifurcation}). 
This implies that, for $\alpha<1$, the shear rate
can be made arbitrarily large by increasing the stress.
In this case, we can calculate the relaxation time in
the large stress limit from Eq. \ref{tau equation2}. Only the region $t\sim1/\dot\gamma$ contributes significantly
to the integral, and for large shear rates this region becomes
arbitrarily small. Thus, we may expand the correlator in powers of $t$ [\cite{hinch}], giving 
\eq{\int_0^\infty \ep{-\dot\gamma t}\phi^2(t)\D t \approx
\int_0^\infty \ep{-\dot\gamma
t}\left(\phi^2(0)+\mathcal{O}(t)\right)\D t.} 
The initial
condition $\phi(0)=1$ (normalisation) gives
\eq{\tau\approx\tau_o+v_2(\sigma)\int_0^\infty \ep{-\sigma t/\tau} \D t}
which is solved in the limit of large stress by
\eq{\tau=\frac{\tau_o}{1-\alpha}\;\;\;\;\;\;(\alpha<1).} 
If $\tau$ grows sublinearly with $\sigma$ at large stress, both quantities
 diverge in this limit. 
In this case
$\alpha\ge 1$ and $\tau=\frac{\tau_o}{1-\alpha}$; demanding a non-negative relaxation time leads to the conclusion that  $\alpha=1$.
Thus, in Model I, $\alpha=1$ is a watershed between ergodic and nonergodic solutions
in the limit of large stress. For $\alpha<1$, solutions are ergodic
with the viscosity approaching a limiting value; 
but for $\alpha>1$, the flow rate $\dot\gamma$ is zero at high stress.

\subsection{Numerical Solutions}
The preceding analytical results are useful but restricted
 to certain model variants. Also they do not provide closed form expressions for flow curves (only transition points). We have therefore also solved our schematic rheological models numerically, using an
established algorithm for MCT models [\cite{Matthias1991}]. The numerical method involves re-writing 
the governing equation in such a way that $\phi(t)$ is determined from
its value at earlier times, allowing iterative calculation of
correlation functions. Numerical efficiency is aided by treating
slowly varying quantities as constant over the timestep.
To adapt this algorithm to our models, we employ the
following procedure:
\begin{enumerate}
\item Input the parameters of interest.
\item Make an initial guess at the relaxation time $\tau$
for these parameters.
\item Calculate the shear rate using this relaxation time and the
chosen value of the stress. This defines iteration-dependent coupling
constants $v_n(\sigma)$ and the memory loss function $f(\dot\gamma t)$.
\item Solve the resulting model via the methods of
 \cite{Matthias1991} (holding the coupling constants fixed).
\item Re-calculate the relaxation time by integrating the resulting correlator.
\item Repeat from step 3 onwards using the new value of the relaxation
time until converged. The result is a self-consistent solution. 
\end{enumerate}

Numerical solutions using this algorithm were obtained using a FORTRAN
code. We checked that the code returned the expected F2, F12 results for $\sigma=0$.
For some parameters close to a nonergodicity transition, the code became slow
to converge, especially for `bad' initial guesses. 
This problem was circumvented by starting the iteration off with a range of initial values and bracketing the correct value according to whether the
initial trajectories showed increasing or decreasing $\tau$.
In some cases this generated improved initial values from which the above algorithm could attain convergence; in others, the process was simply continued until converged to the required accuracy in its own right. (The stability of results obtained in this manner was then
checked by a final iteration.)

\section{Results: Model I}\label{Results}
In this Section we present detailed results for Model I. Later (Section \ref{variations}) we shall examine the effect of altering the form of the memory function, resulting in  Models II--V. 
(This is easier than attempting to present results for all the model variants in parallel; it does not necessarily mean that we prefer Model I to the rest.) From now on we use units where the local
relaxation time far from the glass transition, $\tau_o$, is unity.
We continue to set the transient elastic modulus $G$ to unity and can therefore identify the viscosity $\eta$ with the terminal relaxation time $\tau$.

\subsection{Steady-State Rheology}
\subsubsection{Choice of parameters}
We expect the most interesting
behaviour to arise close to the quiescent stress-free 
glass transition, where there is maximal interplay between the long
relaxation times and the effect of shearing. 
For hard-sphere colloids,
discontinuous thickening tends to occur for concentrations
$\phi\gtrsim 0.5$, for which the zero-shear viscosity is around $50$
times that of the solvent [\cite{MeekerEtAl97}]. 
Thus, we first study
values of $v_2$ for which 
$\tau\gtrsim 50$ (recall that $\tau_o=1$).
The stress-coupling parameter $\alpha$
does not have such an obvious experimental analogue, and it is less
clear what values to investigate. However, from Sec. \ref{solution},
we expect values of order unity to lead to stress-induced
nonergodicity, so we mainly consider such values. 
(As indicated in Sec. \ref{highstress}, and confirmed below, $\alpha = 1$ in fact separates differing regimes
of behaviour.)
Stress can result in
thickening, in principle, whenever $v_2+\alpha\sigma\simeq 4$: but
whether this actually occurs
depends on the interplay with shearing-induced memory loss. 

\subsubsection{Dependence on $v_2$}
\begin{figure}
\begin{center}
\includegraphics[width=80mm]{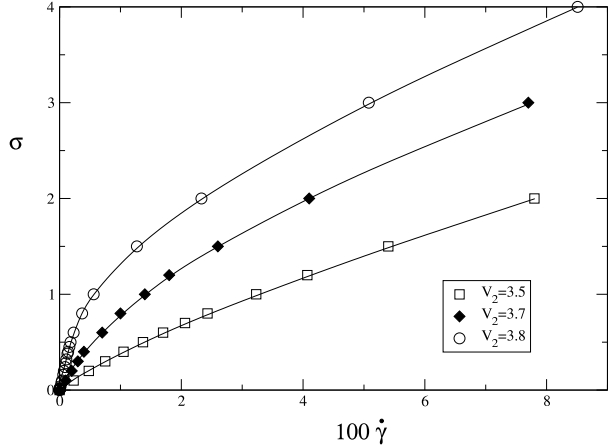}
\caption{Model I flow curves for $\alpha=0.95$: points are numerically
determined values of the shear rate for a given stress.
Shear thinning is seen and, for $v_2=3.8$, thickening is
also discernable at low stresses.}
\label{flowcurves1}
\end{center}
\end{figure}

We begin by varying $v_2$ at a fixed value of
$\alpha$, chosen slightly less than unity. 
Fig. \ref{flowcurves1} shows the flow curves for
$\alpha=0.95$ with $3.5\le v_2\le3.8$. In all cases, there is a regime of shear
thinning and, for the largest value of $v_2$, the shear thinning
is preceeded by thickening at small applied stresses. In all
flow curves presented (here and below) lines are to guide the eye. 
For more modest values of $v_2$ the rheology, not shown,
is close to Newtonian. (For $v_2=3$, for example, the viscosity 
changes only by about $15\%$ upon increasing the stress between zero and
$\sigma=5$.) 

\begin{figure}
\begin{center}
\includegraphics[width=80mm]{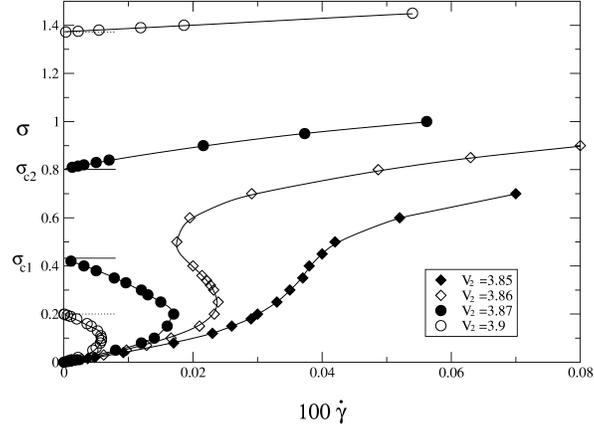}
\caption{Further Model I flow curves for $\alpha=0.95$. For the two largest values of $v_2$, there is a
window of stress for which the viscosity diverges. The limits of this
window (denoted $\sigma_{c1}$ and $\sigma_{c2}$, as shown here for
$v_2=3.87$), are jamming transitions. The predicted values of
these, according to Eq. \ref{transitions}, are
shown as horizontal line segments near the stress axis.}
\label{flowcurves2}
\end{center}
\end{figure}

Fig. \ref{flowcurves2} shows how the trend towards shear thickening 
develops upon further increasing $v_2$. 
In contrast to the rather gentle curves of Fig. \ref{flowcurves1},
the behaviour observed here is rather drastic.
As $v_2$ is increased above $3.8$, the shear thinning
behaviour shown in Fig. \ref{flowcurves1} is interrupted by  shear
thickening whose severity ranges from modest to extreme -- for 
the larger values of $v_2$ shown, the thickening becomes so strong
that regions of negative slope appear. In these regions, increasing
the stress {\it lowers} the shear rate. Such regions of negative slope 
should be mechanically unstable to shear banding [\cite{Dhont,
  OlmstedReview}] and should lead in practice to discontinuous shear thickening as we discuss in Sec. \ref{Discuss}.

For the largest values of $v_2$ shown, there is a window of stress
for which the viscosity has diverged, resulting in ``full jamming'' --
the creation of a nonflowing state by application of stress
[\cite{HolmesEPL,Hess}]. Within this window, the only solution is a
nonergodic, jammed state. The edges of
this window are the jamming transitions whose positions are
given analytically by Eq. \ref{transitions}. The analytic and
numerical results for their positions are in  agreement.
We examine in more detail the
nature of these transitions in Sec. \ref{Discuss}.

The above scenario describes the stress induced arrest of a state with $v_2<4$; this is a liquid at low stresses and therefore the region
of the flow curve close to the origin is always Newtonian (albeit with a viscosity that diverges as $v_2 \to 4$).  By setting $v_2\ge 4$, we can consider shearing a system that is arrested in the quiescent state (the usual definition
of a colloidal glass).
\begin{figure}
\begin{center}
\includegraphics[width=80mm]{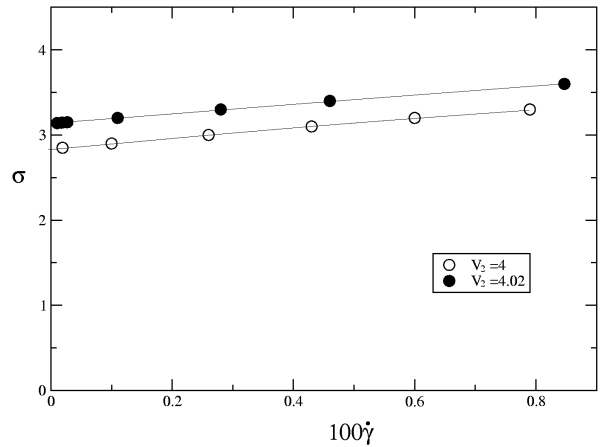}
\caption{Model I Flow curves for $\alpha=0.95$, with $v_2\ge v_2^c=4$. 
Steady state flow occurs only for stresses above the yield stress.}
\label{flowcurves3}
\end{center}
\end{figure}
In this case, as shown in Fig. \ref{flowcurves3}, there is no steady-state flow unless the stress is
increased beyond a yield stress which itself increases monotonically with $v_2$. (No new features arise at larger $v_2$ than those shown here.) 
The application of stresses beyond the yield stress lead to
`shear melting' of the glass, followed by a thinning viscosity, similar
to the behaviour at stresses above $\sigma_{c2}$ in
Fig. \ref{flowcurves2}. 

Our results for Model I with $\alpha=0.95$ can then be
summarised as follows: far from the quiescent glass transition
($v_2\lesssim 3$), the rheology is quasi-Newtonian; for $3\lesssim
v_2\lesssim 3.8$, shear thinning is apparent. In the range $3.8\lesssim
v_2 < 4$, shear thickening becomes prominent: upon
increasing $v_2$ the flow curve first becomes
nonmonotonic, and then (for $v_2\gtrsim 3.866$) the curve reaches
all the way back to the stress axis, indicating a jamming transition
to a nonergodic state. Finally, for $v_2\ge 4$, a conventional 
yield stress appears,
below which there is no (steady-state) flow. At stresses exceeding this yield stress,
behaviour is similar to that above $\sigma_{c2}$ in the
`full jamming' flow curves.

\subsubsection{Dependence on $\alpha$}\label{alpha}
\begin{figure}
\begin{center}
\includegraphics[width=80mm]{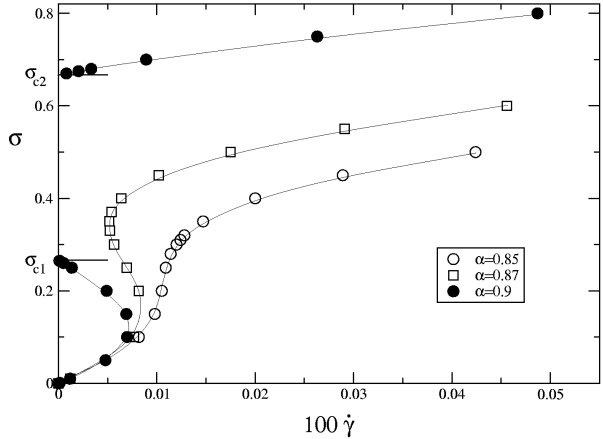}
\caption{Model I flow curves at $v_2=3.9$ for various $\alpha$. For
  $\alpha=0.9$, jamming transitions and their analytical locations
are indicated as in Fig. \ref{flowcurves2}.}
\label{flowcurves4}
\end{center}
\end{figure}

We now fix the value of $v_2$ and investigate the
effect of varying $\alpha$. As mentioned earlier, the
experimental meaning of $\alpha$ is not obvious: it controls the
extent to which the mode coupling vertex increases with applied
stress, and seems related to the susceptibility of the static
structure to external forcing. 
Here we choose $v_2=3.9$ (for which the
zero-shear viscosity is $\tau\approx814$): this is close enough
to the quiescent glass transition to uncover some interesting behaviour.
This behaviour arises for $\alpha$ values of order unity; note that if instead
$\alpha \ll 1$ there is no shear thickening. 
Flow curves for a range of $\alpha\lesssim1$ are shown in Fig. \ref{flowcurves4}.
The variation in the curves with increasing $\alpha$ is qualitatively
similar to
that in Fig. \ref{flowcurves2}, with the shear thickening becoming
stronger as $\alpha$ is increased. 

\begin{figure}
\begin{center}
\includegraphics[width=80mm]{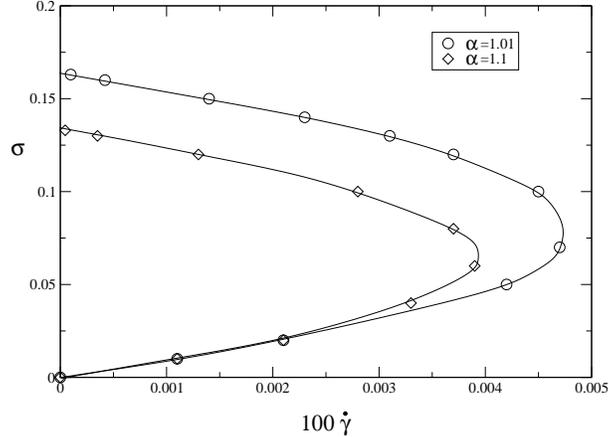}
\caption{Model I flow curves at $v_2=3.9$ with $\alpha > 1$.  There is
  no refluidisation from the jammed state to a high stress fluid.}
\label{flowcurves5}
\end{center}
\end{figure}

\begin{figure}
\begin{center}
\includegraphics[width=80mm]{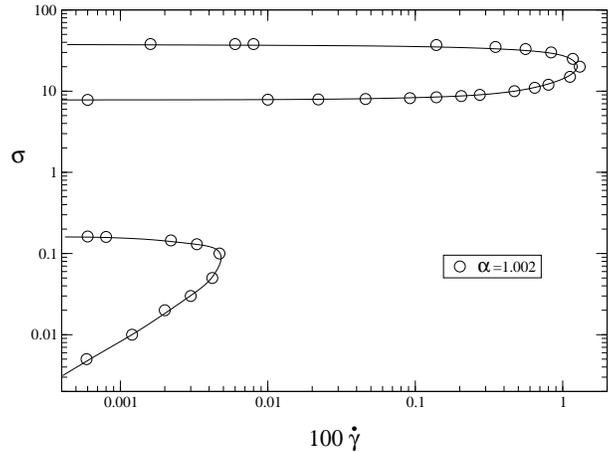}
\caption{Model I flow curve for $\alpha=1.002$ (in a log-log plot),
showing two windows of jamming. Fluidisation from the second
window by increasing the stress is not possible.}
\label{flowcurves6}
\end{center}
\end{figure}

In fact there is an important qualitative distinction between
Fig. \ref{flowcurves4} and Fig. \ref{flowcurves2} which is not
immediately apparent from the plots. Specifically, as $\alpha$ is
varied, the asymptotic slope of the flow curves at large stresses
diverges as $\tau \sim 1/(1-\alpha)$ (Sec. \ref{highstress}). In
contrast, varying $v_2$ at constant $\alpha<1$ leads always to curves of
the same (finite) limiting slope.
For $\alpha > 1$, the slope of the flow curve at large stresses
remains infinite which means there is no upper branch at large
$\dot\gamma$ as illustrated in  Fig. \ref{flowcurves5}.
This is consistent
with our earlier conclusion that, for
$\alpha >1$, $\lim_{\sigma\to\infty}\tau=\infty$. However, the topology of Fig. \ref{flowcurves5} is not
the only possibility: there is a  small  region of parameter space 
($1<\alpha\lesssim 1.0032$ for $v_2=3.9$) for which the system
refluidises upon increasing the stress, 
followed by a {\it second}
window of nonergodicity, which persists indefinitely upon further
increasing $\sigma$. This somewhat peculiar behaviour is
illustrated in Fig. \ref{flowcurves6}.

A remark is in order on the physical meaning of flow curves that show
no liquid-like branch at high stresses. We do not expect that
applying an arbitrarily large stress will fail to make a colloidal
system flow at all; but the flow could be unsteady, or could be
spatially nonuniform involving a fracture mechanism, for
example. Neither outcome is addressable within the modelling framework
developed here, so that flow curves such as those of
Fig. \ref{flowcurves5} need not be regarded as unphysical in our
context.
On the other hand, whenever $\alpha > 1$ such flow curves arise
even for small $v_2$, far from the quiescent glass transition. This 
would surely be unphysical so we assume that if $\alpha$ ever exceeds 1, it does
this only at high concentrations. 
 
A final, intermediate flow scenario is found at the critical
value $\alpha=1$. In this case it
is difficult
to resolve some important details of the flow curve numerically: there
is nontrivial structure in regions where $\sigma$ and $\tau$ are
large. There is certainly a `low stress' window
of nonergodicity for $0.165\lesssim \sigma \lesssim 5.935$ (for $v_2=3.9$). Upon
increasing the stress further, the jammed state yields, followed
by a second regime of thickening in which $\tau$ becomes
too large to track numerically. This suggests that a
second nonergodicity transition might occur: however, we shall see in 
Sec. \ref{phaseDiagrams} (by examining Eq. \ref{transitions}) that
this is not the case.

\subsection{Phase Diagrams}\label{phaseDiagrams} 
The above concludes our survey of flow curves for Model I, which has
covered the basic rheological scenarios found. To get an overview of
the behaviour over a wider range of parameters, we now concentrate
on the jamming transitions and their locations in parameter space.
By solution of Eq. \ref{transitions} for a range of $v_2$ and
$\alpha$, we can calculate `phase diagrams' depicting the 
boundary in parameter space between ergodic and
nonergodic regions. 

\subsubsection{The $(\sigma, \alpha)$ representation}
\begin{figure}
\begin{center}
\includegraphics[width=80mm]{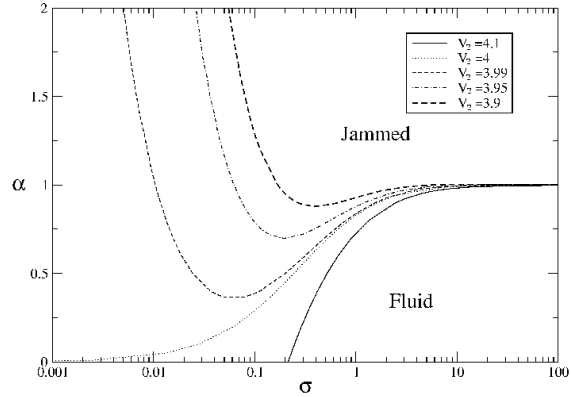}
\caption{Phase diagrams for Model I at various $v_2$. The curves
denote jamming transitions. All states below 
the curve for a given value of $v_2$ are fluid states, whilst those
above (and on) the line are nonergodic, jammed states.}
\label{pdiagram}
\end{center}
\end{figure}

\begin{figure}
\begin{center}
\includegraphics[width=80mm]{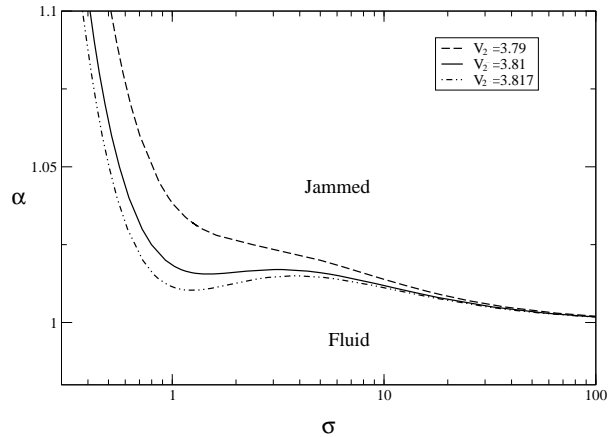}
\caption{Phase diagrams for Model I at smaller $v_2$. (Note the
different scale to Fig. \ref{pdiagram}.) As before, jammed
states lie on or above the line for a given value of $v_2$.}
\label{pdiagram2}
\end{center}
\end{figure}

Phase diagrams of Model I, in the $(\sigma, \alpha)$--plane, are shown
in Fig. \ref{pdiagram} and Fig. \ref{pdiagram2}.
This representation of the phase behaviour shows that, at
a given  value of $v_2$, nonergodic
states appear provided $\alpha$ is sufficiently
large.  As $v_2$ is lowered below $4$ (so that the system
at rest is ergodic), the minimum
value of $\alpha$ necessary for nonergodicity under stress increases from zero until,  for
$v_2$ less than a critical value $\tilde v_2$ (found below) there is no
transition unless $\alpha >1$. Once this condition applies, a transition occurs 
regardless of any further decrease of $v_2$  (Eq. \ref{largeStressTrans}).

In Fig. \ref{flowcurves4} (respectively Fig. \ref{flowcurves5}), we showed examples
of Model I
flow curves in which a jammed state could (could not) be
refluidised by further increasing the stress. Figs. \ref{pdiagram},\ref{pdiagram2} show
that such refluidisation occurs within a window
of $\alpha$ which disappears when $v_2$ falls below a second critical
value $\tilde{\tilde v}_2$ (also found below). Outside of this
window, the upper, refluidised branch of the flow curve is absent (Fig. \ref{flowcurves5}). 
When such a window of $\alpha$ is present at $\alpha>1$ (visible 
for the lower two curves in Fig. \ref{pdiagram2})
an upper branch of the flow curve exists, but its
relaxation
time diverges in the large
stress limit and so this branch shows a second jamming transition with no
ultimate fluidisation at high stress (as in Fig. 
\ref{flowcurves6}). Flow curves showing full jamming with a single refluidised branch (as visible in Figs. \ref{flowcurves2},\ref{flowcurves4}) thus require both $\alpha$ somewhat less than 1 and $v_2$ somewhat less than $4$.
 
\subsubsection{The $(\sigma,V)$ representation}
\begin{figure}
\begin{center}
\includegraphics[width=80mm]{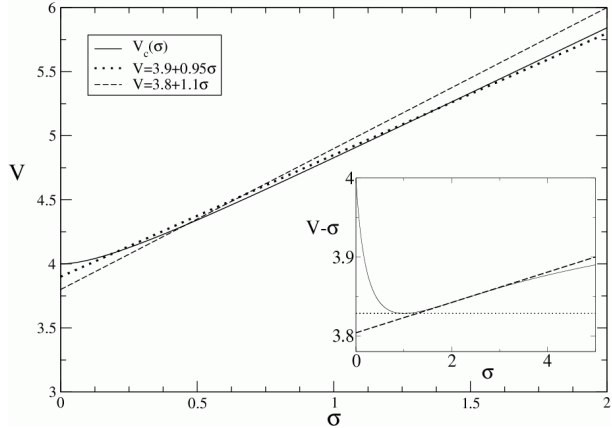}
\caption{Phase diagram in the ($\sigma$, $V$)--plane. Above and on the
  transition boundary, states are nonergodic. The dotted line
  corresponds to $v_2=3.9$, $\alpha=0.95$: the two intersections of
  the transition boundary correspond to jamming, followed by
  refluidisation. The dashed line (corresponding to $v_2=3.8$, $\alpha=1.1$) has a single
  intersection, showing that for these parameters there is no
  refluidisation of the jammed state.  In order to bring out the
  structure present in $V_c(\sigma)$, in the inset we plot
  $V_c(\sigma)-\sigma$ (solid line) as a function of $\sigma$. We show the
  tangent to this curve at its point of maximum slope (dashed line),
  whose intercept with the vertical axis
  identifies $\tilde{\tilde{v}}_2$. (See main text: the geometrical
  interpretation of $\tilde{\tilde{v}}_2$ holds equally well in this
  representation as the subtraction of a linear term
  does not change curvature.) For
  $\alpha=1$ the line $V(\sigma)-\sigma$ is
  independent of $\sigma$ and has the value $v_2$. Thus we identify
  $\tilde{v}_2$ (see text) as the smallest value taken by
  $V_c(\sigma)-\sigma$: the dotted line corresponds to $V-\sigma=\tilde{v}_2$.}
\label{sollichPdiagram}
\end{center}
\end{figure}

For any given parameter set, the $(\sigma, \alpha)$ representation
of the phase diagram
provides a useful picture of the behaviour under increasing the
stress. However, the full range of jamming scenarios is most easily
summarised using a different representation. This involves plotting the locus of
transitions in the ($\sigma, V)$ plane, where we recall that $V\equiv v_2(\sigma)$. We refer to the resulting transition curve
as $V_c(\sigma)$; it is found by solving Eq. \ref{transitions}
for $v_2$ at fixed $\sigma = \sigma_c$, and has no dependence on parameters
such as $\alpha$.
Now, in Model I, $V(\sigma)$ is just a straight
line 
with slope $\alpha$ which intercepts the $V$ axis at $v_2$. Each
transition point, for given $v_2$ and $\alpha$, corresponds to an
intersection of $V(\sigma)$ with $V_c(\sigma)$. The
range of behaviour obtained upon varying $v_2$ and $\alpha$ 
simply corresponds to the different ways in which a straight line
may intersect the fixed locus of possible transition points represented by the curve $V_c(\sigma)$. This is 
illustrated in Fig. \ref{sollichPdiagram}.

\subsubsection{Implications for $\tilde v_2, \tilde{\tilde v}_2$}
Above, we stated that, for $v_2<\tilde{\tilde v}_2$
multiple transitions are no longer allowed. The $(\sigma,V)$ representation
allows a precise calculation of $\tilde{\tilde v}_2$, as follows. From Eq. \ref{transitions}, the
maximum slope of $V_c(\sigma)$ is $V^\prime_c(\sigma_{max})=
27-15\sqrt{3}\approx1.019$ at $\sigma_{max}=1+2/\sqrt3$. For
$\alpha\ge V^\prime_c(\sigma_{max})$, there is then a single jamming
transition if $v_2<4$.
For refluidisation to occur, the jamming transition must occur
at a stress $\sigma <\sigma_{max}$: beyond this point the slope 
$V_c^\prime(\sigma)$ is
monotonically decreasing and so a straight line cutting the transition
curve (from below) beyond $\sigma_{max}$ cannot encounter it subsequently.
The critical value $\tilde{\tilde v}_2$ is then found by setting
$v_2+V^\prime_c(\sigma_{max})\sigma_{max}=V_c(\sigma_{max})$; that is,
$\tilde{\tilde v}_2$ is the
intercept on the $V$ axis of the tangent to the point
of maximum slope.  This gives $\tilde{\tilde v}_2 =
9(1-1/\sqrt{3})\approx 3.804$. This geometrical interpretation 
is illustrated in the inset to Fig. \ref{sollichPdiagram}.

The slope of the transition line $V_c(\sigma)$ approaches unity at large
stresses ensuring that, for $\alpha >1$, the system is nonergodic
in the limit. In contrast, for $\alpha< 1$, jamming transitions
occur only if $v_2$ is sufficiently large. The case $\alpha=1$ is of
particular interest, as we were unable to elucidate its behaviour
numerically. The minimum value of $v_2$ for which a transition occurs
(with $\alpha=1$) is $\tilde v_2$, as defined earlier. 
This is now obtained by asking what is the minimum intercept (with
the $V$--axis) for which a straight line of unit slope cuts the
transition line $V_c(\sigma)$. This 
yields a value of $\tilde v_2 = 1+2\sqrt2\approx 3.828$.  (Again, the
geometrical interpretation, suitably altered, is illustrated in the
inset to Fig. \ref{sollichPdiagram}.) 
At $\alpha=1$, there is no transition below this value of $v_2$; for $\tilde v_2<v_2<4 $ there
are two transitions (jamming and
refluidisation at $\sigma_{c1}$ and $\sigma_{c2}$ respectively). As
$v_2\rightarrow (1+2\sqrt2)^+$, $\sigma_{c1}$ and $\sigma_{c2}$ merge,
and as $v_2\rightarrow 4^-$, $\sigma_{c1}\rightarrow 0$ and
$\sigma_{c2}\rightarrow\infty$. 

We have shown that our critical values of $v_2$ obey $\tilde v_2 >
\tilde{\tilde v}_2$. For $v_2$ between these two values, flow curves
with two separate jammed states (Fig. \ref{flowcurves6}) can be found,
but there are no flow curves with
a single jammed state refluidised at high stress, as seen in
Fig. \ref{flowcurves2} or \ref{flowcurves4}.

\subsubsection{Yielding of a quiescent glass}
Finally, this representation of the phase diagram allows us to
elucidate the yielding of a quiescent glass within Model I. 
As $\sigma\rightarrow 0$, $V_c(\sigma)\rightarrow 4$, and
$V^\prime_c(\sigma)\rightarrow 0$. 
Therefore, for $\alpha=0$, the
yield stress is zero at the quiescent transition ($v_2=4$) and increases
smoothly upon moving deeper into the glass. 
This differs from the microscopic MCT calculations of
\cite{Fuchs+CatesPRL,Fuchs+CatesFaraday} where, although there is no
explicit stress dependence of the vertex, the yield stress rises
discontinuously from zero at the static glass transition. However, in Model I
any $\alpha$ obeying $0<\alpha<1$ restores this expected discontinuity of yield stress
at the static F2 transition.
The resulting yield stress increases with $\alpha$ and diverges for $\alpha\ge1$, beyond which, within Model I, the 
glass cannot be shear melted at any finite stress.

\subsection{Relaxational dynamics}\label{relaxation}
The results presented thus far are determined solely by
the relaxation time (equivalently, viscosity) for given parameter sets. 
This is the time integral of the correlator.
We now consider the functional form, rather than just
the time integral, of the correlator $\phi(t)$ in Model I. 

\begin{figure}
\begin{center}
\includegraphics[width=80mm]{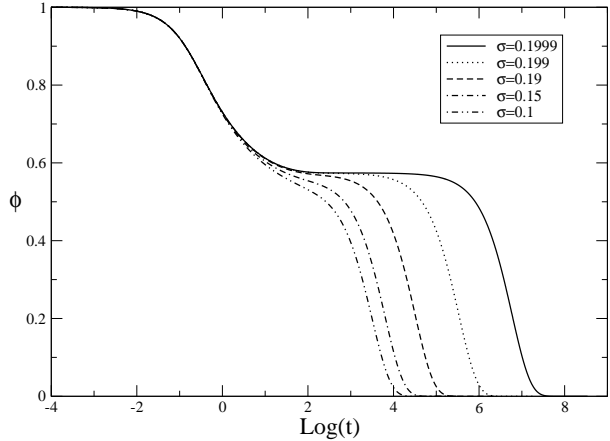}
\caption{Model I correlation functions for stresses approaching
$\sigma_{c1}=0.2$ for $v_2=3.9$, $\alpha=0.95$.
Sufficiently close to the transition, there is a clear separation
between the $\beta$--decay and the flow induced final decay,
resulting in a well defined plateau. At lower stresses, the
relaxation time $\tau$ become comparable to that of the
$\beta$--relaxation, and the plateau is lost.}
\label{correlators1}
\end{center}
\end{figure}

\begin{figure}
\begin{center}
\includegraphics[width=80mm]{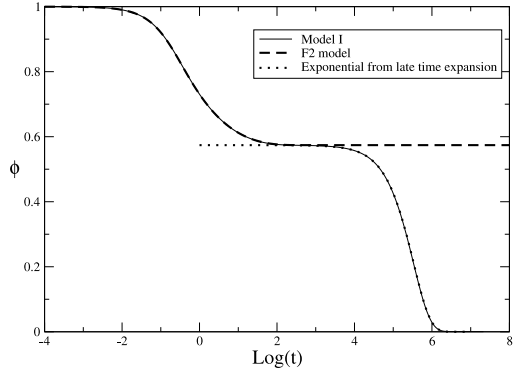}
\caption{The correlation functions for Model I with $v_2=3.9$,
$\alpha=0.95$ and $\sigma=0.199$, compared with the correlation
function for the F2 model with $v_2=3.9+(0.95\times 0.199)=4.08905$,
and (at late times) with the exponential consistent with
Eq. \ref{alpha-decay}. At $\sigma=0.199$, the shear rate is 
$\dot\gamma=1.04 \times10^{-6}$ (the
initial jamming transition for these parameters is at $\sigma_{c1}=0.2$).}
\label{F2comparison}
\end{center}
\end{figure}

Fig. \ref{correlators1} shows the
correlation functions on approaching a typical jamming transition.
At early times the correlators for a range of applied stress are very
similar: the $\beta$--relaxation
timescale (on which the initial decay of correlations occurs) is
approximately constant. The correlators differ significantly only in their
terminal relaxation times
$\tau\sim1/\dot\gamma$.  This shear-induced relaxation time diverges as the
jamming transition is approached ($\dot\gamma\to 0$). 
The consequences of this are as discussed in Section \ref{analytic}. 
On timescales $t\ll 1/\dot\gamma$,
Model I behaves as a static F2 model with coupling
constant $V\equiv v_2+\alpha\sigma$ which, unless $V=4$, 
has a finite $\tau_\beta$. Fig. \ref{F2comparison}
illustrates the equivalence of correlations in Model I, at times $t\ll 1/\dot\gamma$, with the `underlying', noncritical, F2 model.

In calculating the whereabouts of transition points
(Sec. \ref{analytic}) we assumed that, in the vicinity of a jamming
transition, correlation functions were well described by an
exponential at late times, and cited numerical evidence. 
Fig. \ref{F2comparison} demonstrates this for one particular parameter
set. A similar agreement has
been found near jamming transitions for other parameters,
including transitions at both $\sigma_{c1}$ and $\sigma_{c2}$.

\section{Models II--V}\label{variations}
Having studied Model I in some detail, we turn
now to consider results from Models II--V, focusing on qualitative 
differences between these and Model I. By doing so,
we hope to clarify which results depend on particular
choices made in setting up that model, and which are more robust.

\subsection{Model II}
\begin{figure}
\begin{center}
\includegraphics[width=80mm]{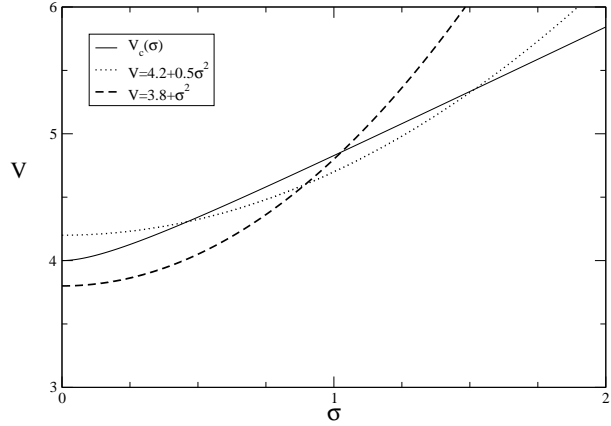}
\caption{The $(\sigma, V)$ representation of the phase diagram for
  Model II. As in Fig. \ref{sollichPdiagram}, intersections of the
  transition line $V_c(\sigma)$ with $V(\sigma)$ mark the jamming
  transitions. The quadratic stress dependence of $V(\sigma)$ in Model
  II ensures at least one jamming transition. At high enough 
stress, Model II is nonergodic, although (as shown) a static glass may
  be shear melted before jamming.}
\label{ModelIIsollichPdiagram}
\end{center}
\end{figure}

The analytic work of Sec. \ref{analytic} applies to Model II, and so
the phase behaviour is again determined via Eq. \ref{transitions}. (As
with Model I, numerics show that the late-time decay of correlations is well
described by an exponential 
verifying the analysis leading to
Eq. \ref{transitions}.) Using the $(\sigma, V)$ representation of the
phase diagram we can elucidate the phase behaviour of this model. 
The transition line $V_c(\sigma)$ remains unchanged
from Model I, whereas $V(\sigma)$ is now quadratically increasing with
stress. Since the maximum slope of $V_c(\sigma)$ is finite, for sufficiently large stresses
$V(\sigma)>V_c(\sigma)$ and so, provided only that 
$\alpha>0$, Model II exhibits a jamming transition
irrespective of other parameters. This ensures that, in the limit of
high stress, Model II is nonergodic. Stress-induced jamming is never followed by refluidisation: and while a quiescent
glass may be shear melted (depending on $\alpha$), this will always
jam again
at higher stresses. These
scenarios are illustrated in Fig. \ref{ModelIIsollichPdiagram}. We
have obtained flow curves (not shown) corresponding to both
of these scenarios. 

\begin{figure}
\begin{center}
\includegraphics[width=80mm]{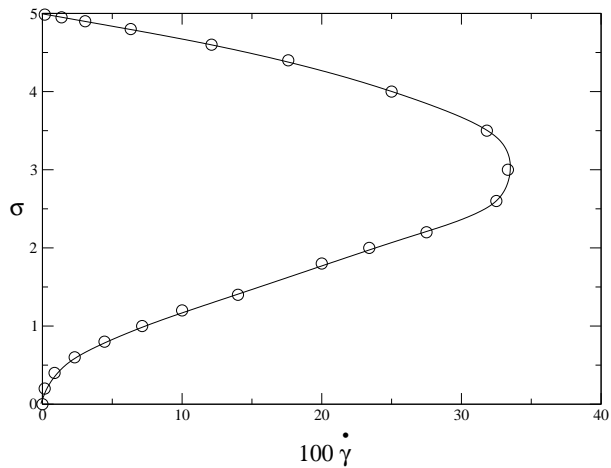}
\caption{Model II flow curves for $v_2=3.9$, $\alpha=0.2$. Note the presence of
thinning at low stresses, followed by thickening and jamming. No
fluidisation after stress-induced jamming is found for these (or
any) parameters with this model.}
\label{extreme}
\end{center}
\end{figure}

The flow curves of Model II also differ from Model I at low stresses
for a system that is liquid at rest; here one finds (for suitable
parameters) a region of downward curvature near the origin. Hence
there is an initial shear thinning region before the shear thickening
and arrest scenarios are encountered. 
Fig. \ref{extreme} shows the flow curves for Model II with $v_2=3.9$,
$\alpha=0.2$, for which full jamming is present: compare this to the
flow curves for $v_2=3.9$ in Model I (Fig. \ref{flowcurves4}), in
which no thinning is apparent prior to thickening. 

\subsection{Model III}
We have just seen that in Model II the system always jams at high stresses but is shear thinning at low ones. In Sec. \ref{model} we argued
that, around $\sigma=0$, the quadratic form (Model II) has a sounder
basis ($v_2(\sigma)$ should be a symmetric function) whereas there is no reason to restrict attention to a quadratic at
larger stresses. The wider variety of flow scenarios emerging from Model I at
high stressses suggests some utility for a model in which a quadratic
stress dependence at low stresses goes over to a less severe form at
higher ones. Model III (see Eq. \ref{ModelIII}) provides a simple way to allow for this.

\begin{figure}
\begin{center}
\includegraphics[width=80mm]{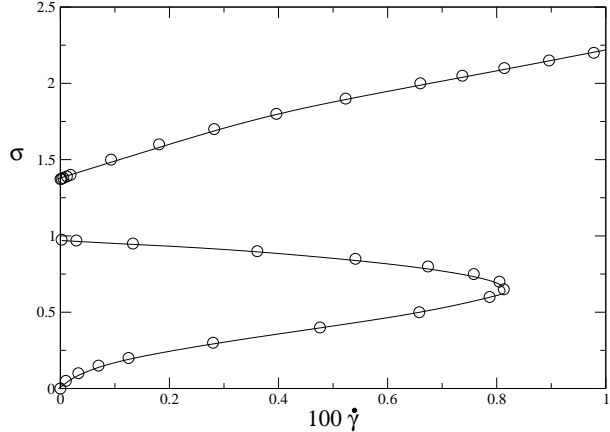}
\caption{The Model III flow curve with $v_2=3.9$, $\alpha=0.95$. Shear
thinning preceeds thickening and jamming.}
\label{quadLinear}
\end{center}
\end{figure}

\begin{figure}
\begin{center}
\includegraphics[width=80mm]{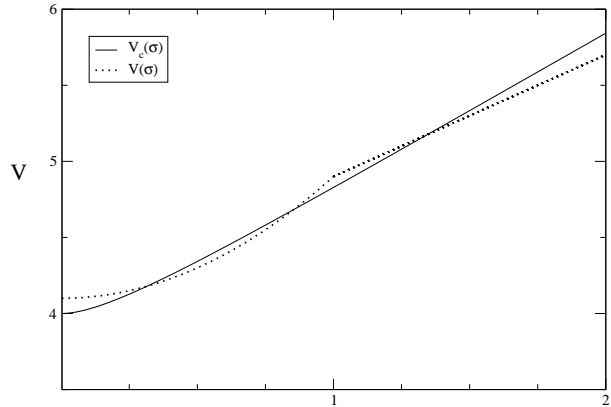}
\caption{The Model III phase diagram in the $(\sigma, V)$
  representation for $v_2=4.1$, $\alpha=0.8$. A shear
  melted glass (or quiescent fluid; not shown) can exhibit thickening and full jamming at larger
  stress, as in Model II (Fig. \ref{ModelIIsollichPdiagram}). But in
  this case, the ultimate state at high stress is a fluid
  for $\alpha < 1$. }
\label{ModelIIIpDiagram}
\end{center}
\end{figure}

As shown in Fig. \ref{quadLinear}, in Model III 
shear thinning can preceed 
thickening, just as in Model II. Unlike
 Model II however, full jamming may now be followed by refluidisation: 
the high stress limit is ergodic if $\alpha$ is smaller
than unity. This is illustrated by the phase diagram trajectory shown
in Fig. \ref{ModelIIIpDiagram}, in which a quiescent 
glass is first fluidised, then jams, and is finally 
refluidised again. 

\subsection{Model IV}
Model IV incorporates a nonexponential form of $f(\dot\gamma t)$
inspired by the full MCT calculations 
[\cite{Fuchs+CatesPRL}].  This has not been treated analytically, and so results
were obtained solely by numerical means. We find that the rheology of
this model is qualitatively similar to that of Model I: the same shear 
thickening scenarios appear, and so we do not show
any flow curves.

\begin{figure}
\begin{center}
\includegraphics[width=80mm]{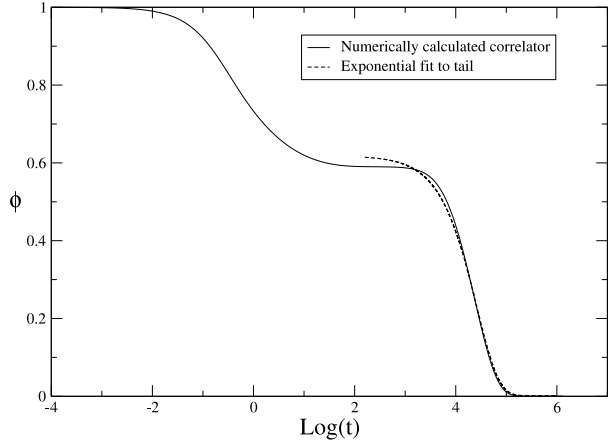}
\caption{The Model IV correlation function for $v_2=3.9$, $\alpha=0.3$
and $\sigma=0.78$ (the system has been refluidised: 
$\sigma_{c2}\approx 0.775$), with an exponential fit to the late-time
decay. The late-time decay deviates from exponential form.}
\label{lateFitAlgebraic}
\end{center}
\end{figure}

\begin{figure}
\begin{center}
\includegraphics[width=80mm]{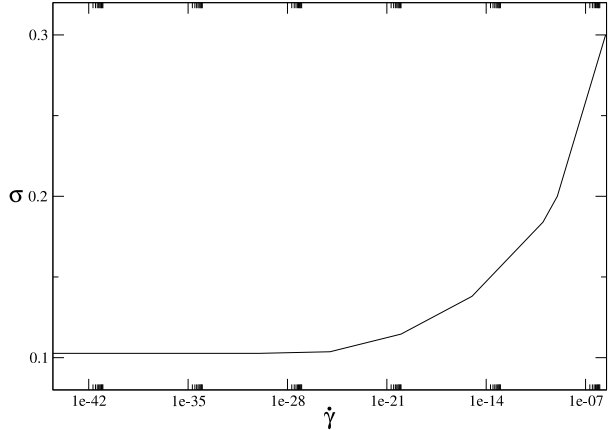}
\caption{The $\alpha=0$ flow curve of Model IV at $v_2=4$. Note the
 logarithmic scale of the $\dot\gamma$ axis. At sufficiently
 small shear rates, the flow curve levels off, and is constant over
 fourteen or so decades in the shear rate, suggesting that
 in this case the yield stress does appear discontinuously at the
 glass transition. Whilst the shear rates required to capture this
 behaviour are exceedingly small, the yield stress itself is not.}
\label{algebraTransition}
\end{center}
\end{figure}

In contrast, the correlation functions {\it do} differ from those
of Model I. In particular, the late-time exponential decay present in Models I
and II does not consistently appear in this case:
in 
Fig. \ref{lateFitAlgebraic} we show an example where the late-time correlator 
 is not well described by an exponential.
This suggests that the exponential decay of Models I, II
is an artefact of their precise form (nevertheless, it is helpful as
it allows analytic progress).
A second difference regards the behaviour in the case $\alpha=0$. In
Sec. \ref{Results}, we showed that, for Model I, the $\alpha=0$ yield
stress vanishes smoothly as $v_2\rightarrow4^+$. Here, elucidating
this point is not so simple, as we have no analytic expression for the
transition points. However, numerics suggest that the yield
stress does {\it not} vanish smoothly as $v_2\rightarrow4^+$ for Model
IV (Fig. \ref{algebraTransition}). We have no explanation for the very
small flow rates needed to reveal this effect.

\subsection{Model V}
We turn finally to Model V, which differs from Model I in that it is based on the F12 
schematic model, rather than the simpler F2 model. 
This creates an extra dimension in the parameter space, since there
are two coupling parameters whose variation with applied stress
need not be identical -- {\it ie}, in Eq. \ref{ModelIV},
$\alpha_1$ need not be identical to $\alpha_2$. 
Our exploration of the parameter space has uncovered thickening and
jamming scenarios qualitatively
similar to that of Models I and IV. Other scenarios may be present, but we
have not found any which differ qualitatively from those shown
earlier.

However, as in Model IV, the correlation functions do not
always show a late-time exponential decay in the vicinity
of jamming transitions. An unresolved issue concerns nature
of the $\alpha=0$ yield
stress: does it appear discontinuously at the quiescent
glass transition (as in Model IV), or does it vanish smoothly (Models
I, II and III)? We have been unable to answer this question numerically:
but in contrast to Model IV, the yield stress, if it exists at all, 
is very small ($\lesssim
10^{-5}$).

\subsection{Structural stability}\label{stability}
Having described the results of the various model variants, we now discuss the
structural stability of our class of models.
The shear thickening/jamming scenarios that we find appear fairly robust with
respect to the precise form of the memory function, indicating that
they arise generically from the competition between stress induced
jamming and flow induced fluidisation, rather than from the precise
mathematical forms chosen here. The exception to this is the
case of Model II (and, more generally, any case in which the coupling
parameters increase too quickly at large stress) in which fluid states 
are precluded at high stress 
regardless of proximity to the quiescent glass transition. This is
unphysical: very dilute colloids do not jam under flow. However, Model II is still useful in that it takes the `correct'  form  around $\sigma=0$, suggesting that
thinning can preceed shear thickening. Model III successfully combines this
with the desirable features of Model I, in which fluid states at high stress
are allowed, but only for $\alpha < 1$. Since $\alpha$ (like $v_2$) can be concentration dependent, this allows an intriguing possibility, that shear melting of colloidal glasses (stress induced or otherwise) is precluded above a certain threshold of concentration.

In contrast, neither the exponential decay nor the smoothly appearing
yield stress at $\alpha=0$ found in Models I -- III
are robust under alterations to the chosen form of the memory
function. In conventional MCT, late-time
decays in the vicinity
of the glass transition are not exponential, but rather are
well described  by the stretched exponential form ($\ep{-(t/\tau)^\beta}$
with the exponent $\beta<1$). This `dynamical stretching', which is found
experimentally, is absent in the F2 model, although it reappears in F12
and other, more general,
schematic models. It is not surprising, then, that Model I
leads to exponential decay whilst Model IV does not. Dynamical stretching
is associated
with the co-operative motion required for relaxation in a dense
liquid. In the vicinity of a jamming transition, relaxation occurs via
smaller scale motions which are then accentuated by the flow. Thus,
although our results regarding this point are inconclusive, one
might expect that dynamical stretching is less pronounced
near these transitions than near the quiescent glass transition
[\cite{Fuchs+CatesFaraday}]. If so, the assumption of an exponential
decay could
be a harmless one.

\section{Discussion}\label{Discuss}
\subsection{The jamming transitions}
Nonergodic solutions to our schematic MCT equations are always
available for $V\ge 4$ (Eq. \ref{bifurcation}).
However, they are not the only
possible solutions. We have seen that ergodic, flowing solutions
can also exist within this region: if both are present,
the iteration method we used for $\tau$ (based {\it eg} on Eq. \ref{tauagain})
leads always to the ergodic solution, and jamming
transitions occur at points where these solutions cease to exist.
So far, we have tacitly assumed that physical stability follows from
such stability under iteration: although plausible, this is not
obvious. We discuss this assumption below, but firstly, presuming that
it holds, we compare our jamming transitions to the conventional
static glass transition as described by standard MCT.

\begin{figure}
\begin{center}
\includegraphics[width=100mm]{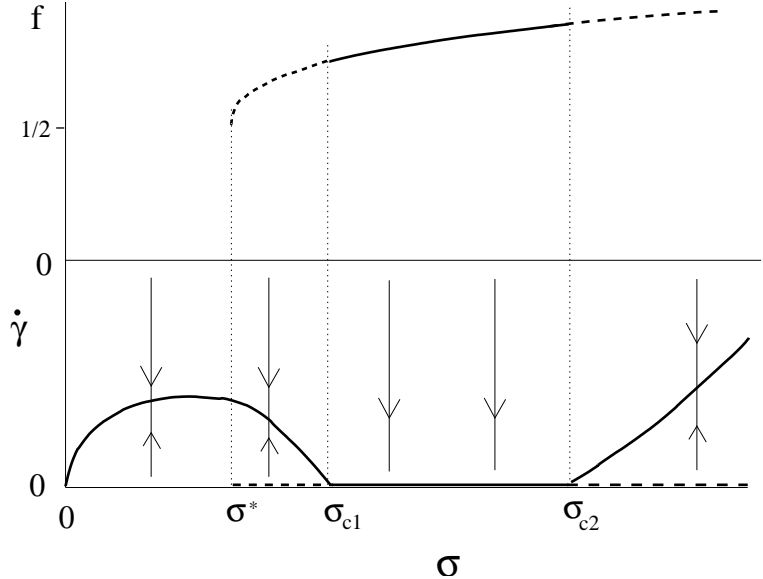}
\caption{Schematic variation of $\dot\gamma$ and $f$ with $\sigma$ (Model I, $\alpha<1$, 
full jamming scenario). Solid and dashed lines denote stable and unstable
solutions under the iteration described in Sec. \ref{solution}. 
Arrows indicate flow of $\dot\gamma$ under the iteration. Note that the ergodic solution remains
stable beyond the critical point $\sigma^\ast=(4-v_2)/\alpha$ where a nonergodic solution first
exists, {\it ie}, $v_2+\alpha\sigma >4$.}
\label{schematic}
\end{center}
\end{figure}

In standard MCT,  glass transitions occur at points
in parameter space where nonergodic solutions become available. At
these so-called `critical points'  [\cite{Goetze}], there is a bifurcation 
in the long-time limit ($f$) of the correlation function:
 it jumps from its value in the
fluid ($f=0$) to a nonzero value $f_c$, followed by a nonanalytic
variation as the coupling is further increased
($f-f_c\sim\epsilon^{1/2}$, where $\epsilon$ is the separation from
the transition in the space of coupling constants). At
lower values of the coupling, no such nonergodic solution is possible.
In the
case of our jamming transitions, this is not the case: rather (at
fixed $v_2$, $\alpha\ldots$) the
system jams at values of the stress for which the ergodic, flowing
solutions vanish. These points are not critical points, and do not
correspond to a bifurcation in $f$. Therefore, the initial jump in $f$
at jamming transitions is followed by an analytic variation with
coupling ({\it ie}, with the stress if other parameters are
fixed). 
This point is illustrated by the variation of the
nonergodicity parameter $f$ with $\sigma$, as shown in 
Fig. \ref{schematic}, which corresponds to one of the `full jamming' flow curves
in Fig. \ref{flowcurves2} or \ref{flowcurves4} for Model I.

In the figure, there is a range of stress for which an arrested solution
has become available (indicated by the existence of nonzero $f$ in the
upper part of the diagram) but is not chosen -- rather, the system
remains flowing (as indicated in the lower part by a nonzero shear
rate). Only when the viscosity of the flowing system diverges (at
$\sigma_{c1}$), signalling that flowing solutions are no longer
possible, does the system become nonergodic. At $\sigma_{c2}$, a
flowing solution again becomes available.

The fact that jamming transitions are not
critical points was used in Sec. \ref{solution} to guide the analysis presented there. 
As discussed also in Sec. \ref{relaxation}, the $\beta$--relaxation 
time $\tau_\beta$, and not just the longer $\alpha$--relaxation,
diverges at a critical point in MCT, whereas at a generic
jamming transition $\tau_\beta$ remains finite. 
This means that, by moving close enough to a jamming
transition, we can make the shear rate sufficiently small that
$\dot\gamma \tau_\beta\ll 1$ so that the underlying
$\beta$--relaxation is essentially unperturbed by the flow. 
This gives a
a pre-plateau relaxation of the correlator on time scale $\tau_\beta$,
followed at time $\tau\sim1/\dot\gamma$ by the flow induced terminal relaxation. 
This leads to a very well defined plateau region, visible in
Fig. \ref{correlators1}, which is quite different from a quiescent MCT
scenario.

This prediction is testable in principle, although the advection of density fluctuations
makes the measurement of correlators somewhat tricky. One idea is to monitor the decay of 
density fluctuations with wavevector along the neutral (vorticity) direction. These fluctuations
are not advected by the flow, so that there is some chance of observing the
characteristic properties of the correlators just described, for example at
stresses just above $\sigma_{c2}$.

\subsection{Physical stability of solutions}
We have so far assumed that stability of solutions to our MCT equations under iteration 
({\it eg} of
Eq. \ref{tauagain}) is equivalent to physical stability of those solutions. There are two
ways in which this assumption can fail. One is well understood, and relates to the mechanical instability of flow curves of negative slope with respect to shear banding. We discuss this
below (Sec. \ref{banding}), but first address a different question -- that of selection between 
different solutions of the schematic MCT equations when more than one exists.

\subsubsection{Iterative versus physical stability}
\begin{figure}[t]
\begin{center}
\includegraphics[width=80mm]{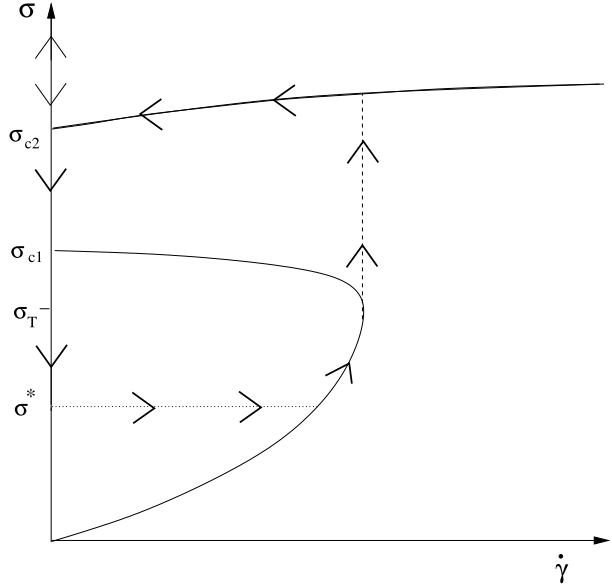}
\caption{Full jamming scenario of Model I, showing in bold the full manifold of
steady state MCT solutions, including nonergodic solutions that are iteratively unstable
(these lie on the vertical axis, outside the window $\sigma_{c1} < \sigma < \sigma_{c2}$).
Dotted line: see text. Vertical dashed line: extreme 
possible discontinuous shear-thickening curve
arising from shear-banding. Bold arrows: extreme possible hysteresis
loop. 
Light arrows: possible exploration of further jammed states.
}
\label{hysteresis}
\end{center}
\end{figure}

As mentioned above, iterative stability selects ergodic solutions over nonergodic ones, 
when both exist. Although such iteration does not
directly map onto temporal evolution of the shear rate,
one could imagine a transient violation of
Eq. \ref{ViscosityPrescription}, resulting in an infinitesimal shear
rate arising in an otherwise nonergodic state. Assuming a correspondence between
iteration and time evolution, this would carry the system to the flowing solution
whenever it exists; the arrows in Fig. \ref{schematic} would be genuine trajectories.
Without knowing this correspondence, however, it seems wise to allow that when nonergodic 
and flowing solutions of our models both exist mathematically, either or both could
be locally stable.  For the full jamming scenario of Fig. \ref{schematic} and Figs.
\ref{flowcurves2}, \ref{flowcurves4}, this leads to the manifold of steady state solutions
depicted by the bold line in Fig. \ref{hysteresis}.

A pessimistic view, on the other hand, is that the nonergodic solution is the only 
physically stable one whenever it does exist (so that physical stability is the opposite
of iterative stability). This assumption would destroy much of the interesting rheology we have
reported; for example, the full jamming scenario seen in Fig. \ref{hysteresis}
would be replaced by a flow curve that consists of the lowest branch of the S-shaped flow
curve for stresses less than $\sigma^\ast$, then jumps discontinuously back to 
$\dot\gamma = 0$ (dotted line) and remains there for all $\sigma > \sigma^\ast$ (vertical
bold line).
This scenario cannot really explain either continuous
or discontinuous shear thickening. Moreover, 
its presumption that a nonergodic state is always preferred to flowing alternative 
contradicts the usual notion that glasses are arrested because they are trapped. 

\subsubsection{Mechanical instability and hysteresis}\label{banding}
Let us assume instead that the full manifold of steady-state solutions are candidates for
physical stability. These solutions describe putatitive steady, homogeneous states; however 
it is well known that any flow curve containing regions of negative slope is 
unstable to shear banding [\cite{OlmstedReview,Dhont}]. For shear thickening systems the bands
comprise slabs of material oriented layerwise in planes perpendicular to the vorticity 
(neutral) axis; these slabs have a common shear rate but different
stresses. (The shear thinning case, in which layers are oriented
parallel to the sliding walls of a couette, may be
more familiar [\cite{OlmstedReview, Spenley}].) In a rheometrical experiment with solid walls,
these stresses get averaged over the slabs; thus, even a stress-controlled experiment controls
only this average.

Accordingly the part of the solution manifold in Fig. \ref{hysteresis} between $\sigma_{c1}$ on the
vertical axis and the point of vertical tangency to the flow curve (at stress $\sigma_T$) is mechanically unstable and will not be observed in steady state, even in a controlled-stress rheometer. The standard shear banding scenario then predicts that if a quiescent
fluid is progressively sheared with increasing $\dot\gamma$, at some point before the 
vertical tangent is reached the stress will jump onto the upper branch of the flow curve.
This is an extreme limit of stability; in practice it could jump sooner, at a shear rate that could depend on various aspects of sample history. (Such phenomena are well studied in the
shear-thinning case; see \cite{Grand97}.) Thus, shear banding explains discontinuous shear
thickening as the natural consequence of an S-shaped flow curve. If $\dot\gamma$ is increased further, the upper branch will be followed out to the
right.

If $\dot\gamma$ is now slowly reduced, there can be hysteresis, in which the jump back to the 
lower branch is delayed. For an S-shaped curve that does not cross the axis (see Fig. \ref{flowcurves2} for examples), this must occur at or before a second point of vertical
tangency, and the hysteresis loop is of limited width (as reported experimentally
in [\cite{Frith,Laun}]). However, in the `full jamming' curve of Fig. \ref{hysteresis} it becomes possible in principle to track right back to
the vertical axis, achieving the jammed state of zero shear rate but a finite stress. 
This offers one way to access the jamming transition at $\sigma_{c2}$ and the
stress-induced glass at $\sigma \le \sigma_{c2}$. 
Another is to perform
a stress-controlled measurement in the window between $\sigma_{c1}$ and $\sigma_{c2}$;
however, since the controlled stress is a spatial average, 
this might lead to a banded flow instead (corresponding, {\it eg}, to a point on the vertical dashed line in Fig. 
\ref{hysteresis}). Once the vertical axis has been reached, it should be possible to 
vary the static stress, at least within the window of full
jamming. It should also be possible to see whether the
nonergodic states outside this window are locally stable; if they are, the stress
could be increased indefinitely (in principle!) without steady flow arising, and/or reduced
as far as $\sigma^\ast$. The latter sets the lower left corner of an extreme
possible hysteresis loop, as indicated in Fig. \ref{hysteresis}.
Note that we have entirely excluded creep from
these discussions which address only steady-state flow and its absence; 
in practice this could complicate matters considerably.

Note also that if a model is chosen with no upper branch to the flow
curve, as in Fig. \ref{flowcurves5} (Model I with $\alpha > 1$) 
the preceding discussion of hysteresis might still apply so long
as such a branch is furnished by some other mechanism, such as fracture. 
Experimentally, it often reported
than flow in the strong shear thickening
regime is far from steady [\cite{Frith}];
this might indeed be consistent with a flow curve such as
Fig. \ref{flowcurves5}, plus an unsteady fracture mechanism at high
stresses. This hybrid
mechanism would allow a complete
separation of stress scales between the upper (shear-thickened) part of the effective
flow curve and the lower (quasi-Newtonian) part; 
in particular, the upper stress level could vastly exceed the 
intrinsic modulus scale for a colloidal glass. The latter 
is of order $k_BT/a^3$ close to the arrest transition, which is of order  1--100 Pa
for typical parameters and decreasing with particle size. Much higher stresses
are indeed sometimes reported in jammed colloids [\cite{Frith, d'Haene}].

\subsection{Comparison with experimental jamming transitions}
As discussed in the Introduction, phenomena reminiscent of our jamming
transitions have been observed experimentally. 
In the work of \cite{Bertrand2002}, jamming
occurred only for a particular range of concentrations: at lower volume
fractions, shear thickening was observed, whilst at higher
concentrations, the sample was solid in its quiescent state.  
A similar sequence arises in our Model I if $v_2$ increases with
concentration (as it surely does) and $\alpha$ remains less than
unity. Thus it is tempting to identify the jamming transitions of our 
model with the stress-induced solids recently 
found experimentally [\cite{Bertrand2002, HawPRL04}]. 

However, such an identification is uncertain. The
particles used in the experiments of \cite{Bertrand2002} 
are perhaps too large for glass transition concepts
to be relevant; they have radius $a\sim 3
\mu$m, so that Brownian forces are small compared with 
forces such as gravity (no
density matching was done) and a purely mechanical description
might be more appropriate. The particles which most easily jam in the
experiments of 
 \cite{HawPRL04} are also at the upper end of the
colloidal size range, with radii $a\sim1\mu$m; it would be
of great interest to study shear-thickening with particle sizes
smaller than this. Also, most experiments on dense suspension rheology
do not utilise the clean model systems latterly favoured by colloid
physicists [\cite{HawPRL04, PhamScience}].
It is possible to closely approximate hard-sphere interactions (this 
generally requires optical index matching with the solvent) to which
attractive forces can then be added controllably. Further
investigations of shear thickening in such model systems would be particularly 
welcome. Ideally one would like to know the behaviour as a function
of concentration, particle size, and interactions. One could then aim
to establish a rational correlation between these and
parameters such as $v_2$ and $\alpha$ in our models.

\section{Conclusions}\label{conclusions}
The rheology of concentrated suspensions is sufficiently complex that
a full microscopic theory is not yet feasible. 
Deliberately simplified models can prove useful in improving our
understanding of these materials. In this article we have presented
a family of models of the rheological behaviour of very dense
suspensions, motivated by the Mode-Coupling Theory of the glass
transition, but neglecting hydrodynamics.

Upon variation of model parameters, a  wide range of rheological
behaviour was found, ranging from shear thinning,
through continuous and discontinuous thickening (the latter interpreted
as shear-banding), to a jamming
transition into a nonergodic solid state. Depending on model parameters,
this may or may not be refluidised again at higher stresses;
in cases where the model predicts no refluidisation one might
expect another mechanism, such as fracture and/or unsteady flow, to intervene.
(There is some experimental evidence for this at very high concentrations.)
Various hysteresis scenarios involving jammed states and/or shear banding 
are also possible.

The main parameters
in our simplest model variant (Model I) are $v_2$, which
is a measure of the distance from the glass transition in
the quiescent state (this lies at $v_2 = 4$), and a coupling
constant $\alpha$ which controls, in effect, the stress dependence of 
this distance.
There is also some freedom in how we model the role of strain
in eroding memory, but these details are less important to the results.

The jamming transitions within our models differ from static glass
transitions, in that the transition points are not
critical points of our schematic MCT: the $\beta$--relaxation time does not diverge at
a jamming transition. The resulting variation of the non-ergodicity parameter 
$f$ at jamming has a discontinuity but is analytic beyond that, unlike
the MCT glass transition which has a square-root contribution. 
Finally, in Models I--III, 
the late time decay of correlators close to a
jamming transition is exponential. However, as discussed
in Sec. \ref{stability}, this is not robust but depends on the
particular form of these models. 
Regardless of this, in all the scenarios suggested in this work, we
expect that this final relaxation should differ from that close
to a conventional glass transition, since the two relaxations occur in
quite different ways.

Although the models presented for shear thickening and jamming remain
both schematic and provisional, they may help guide future progress
with more formal MCT-based developments [\cite{Fuchs+CatesPRL,Fuchs+CatesFaraday}].
We also hope that this work will promote a more careful experimental
examination of the flow behaviour of very concentrated suspensions, preferably
in systems of small, density-matched particles (so that Brownian motion is strong
and gravity weak) with
well-controlled colloidal interactions. The possible interplay between the jamming
we describe and capillary forces at the surface of a droplet, or at 
the fracture surface of a bulk sample ruptured by shear, remains an open
topic for future study. In the longer term, it would be very useful to develop
an MCT-like theory, even at the schematic level, that can address time-dependent
phenomena as well as the steady-state properties addressed in this work.

\section{Acknowledgements}
We thank Patrick Warren, Mark Haw, J\'{e}r\^{o}me
Bibette, Norman Wagner and
Thomas Voigtmann for useful discussions. This work was supported by
the EPSRC, through Grant GR/S10377/01, and by the Deutsche
Forschungsgemeinschaft through Grant Fu309/3.

\bibliography{bibliography}

\end{document}